\documentclass[12pt, a4paper]{article}
\usepackage{graphicx} 
\usepackage{graphics}
\usepackage{fancybox}
\usepackage{amsmath,latexsym,psfrag,epsf,epsfig,amssymb,rotating}
\usepackage{color}
\usepackage{multirow}
\usepackage{url}
\usepackage{natbib}
\usepackage{fullpage}
\usepackage{subfigure}
\usepackage{bm}

\usepackage{dsfont}

\usepackage{algorithm}
\usepackage{algorithmic}

\usepackage[
            colorlinks]{hyperref}  
\hypersetup{linkcolor=blue,
            citecolor=blue,
            filecolor=black,
            urlcolor=blue,
            bookmarks=true
           }
\hypersetup{
    bookmarks=true,          
    unicode=false,           
    pdftoolbar=true,         
    pdfmenubar=true,         
    pdffitwindow=false,      
    pdfauthor={Alberto Caimo, Isabella Gollini},     
    pdfnewwindow=true,       
    colorlinks=true,         
    linkcolor=blue,          
    citecolor=blue,          
    filecolor=blue,          
    urlcolor=blue            
}
\newcommand{\pref}[1]{%
    \ref{#1} \ifnum\count0=\pageref{#1}\relax%
    \else (page \pageref{#1})\fi}

\newcommand{\eref}[1]{%
        \ref{#1}\ifnum\count0=\pageref{#1}\relax%
        \else {, p.\pageref{#1}}\fi}

\newcommand{\comment}[1]{}

\newcommand{\bfxi}{\boldsymbol{\beta}}

\newcommand{\bfphi}{\boldsymbol{\phi}}

\newcommand{\bfLambda}{\boldsymbol{\Lambda}}

\newcommand{\bftheta}{\boldsymbol{\theta}}

\newcommand{\bfeta}{\boldsymbol{\eta}}
\newcommand{\bfgamma}{\boldsymbol{\gamma}}

\newcommand{\bfSigma}{\boldsymbol{\Sigma}}

\newcommand{\bfmu}{\boldsymbol{\mu}}

\def\bfS{\textbf{\em S}}

\def\bfy{\textbf{\em y}}
\def\bfY{\textbf{\em Y}}

\usepackage{authblk}

\usepackage[english]{babel}
\usepackage[utf8x]{inputenc}
\usepackage{amsmath}
\usepackage{graphicx}

\def\nstats{r}

\providecommand{\keywords}[1]
{
  \small	
  \textbf{\textit{Keywords---}} #1
}

\title{A multilayer exponential random graph modelling\\ approach for weighted networks}

\author[1]{Alberto Caimo}
\author[2]{Isabella Gollini}
\affil[1]{\small Technological University Dublin, Ireland; \texttt{alberto.caimo@dit.ie}}
\affil[2]{\small University College Dublin, Ireland; \texttt{isabella.gollini@ucd.ie}}

\begin{document}
\maketitle

\begin{abstract}
A new modelling approach for the analysis of weighted networks with ordinal/polytomous dyadic values is introduced. Specifically, it is proposed to model the weighted network connectivity structure using a hierarchical multilayer exponential random graph model (ERGM) generative process where each network layer represents a different ordinal dyadic category. The network layers are assumed to be generated by an ERGM process conditional on their closest lower network layers. A crucial advantage of the proposed method is the possibility of adopting the binary network statistics specification to describe both the between-layer and across-layer network processes and thus facilitating the interpretation of the parameter estimates associated to the network effects included in the model. The Bayesian approach provides a natural way to quantify the uncertainty associated to the model parameters. From a computational point of view, an extension of the approximate exchange algorithm is proposed to sample from the doubly-intractable parameter posterior distribution. A simulation study is carried out on artificial data and applications of the  methodology are illustrated on well-known datasets. Finally, a goodness-of-fit diagnostic procedure for model assessment is proposed. 
\end{abstract}

\keywords{Statistical network models; Bayesian analysis; Weighted networks; Intractable models.}

\section{Introduction}

Statistical network analysis concerns modelling relationships defined by edges between nodes \citep{tow:whi:gol:mur12}.
In many empirical contexts these relationships have a strength associated with their edges \citep{bar:bar:pas:ves04}. The nature of the variation in the strength of an edge between two nodes may be determined by a variety of aspects depending on the application context; for example, the amount of traffic flowing along connections in transportation networks \citep{ops:col:pan:ram08}, the functional connectivity levels in brain networks \citep{bul:spo09}, and interactions in cellular and genetic networks \citep{hor11}.

In the context of social network analysis, one of the most important families of models used to describe the connectivity network structure is represented by exponential random graph models (ERGMs) \citep{hol:lei81,str:ike90}. These are generative models for networks postulating an exponential family over the sample space of networks on the fixed set of nodes and are specified by a set of sufficient network statistics \citep{bes74} representing sub-graph configurations that are believed to be important to the generative process of the observed network structure. ERGMs are very flexible models as they can potentially incorporate any type of network statistic and were originally defined for networks with binary edges encoding the presence or absence of an edge between two nodes.  Commonly used network statistics include summaries of density (e.g., number of edges), homophily (e.g., number of edges among nodes with the same nodal attribute), degree-based statistics (e.g., number of k-stars), and triad-based statistics (e.g., number of triangles) \citep{sni:pat:rob:han06}.

Several modelling extensions based on ERGMs have been proposed. These include generalisations of binary ERGMs to polytomous forms of weighted edges \citep{robins1999logit, pattison1999logit}; the multi-valued curved ERGMs \citep{wya:cho:bil10}; ERGMs for inference on networks with continuous edge values \citep{des:cra12, wil:etal17}; the Geometric/Poisson reference ERGMs for ordinal/count networks \citep{kri12}; and the rank-order-edge ERGMs \citep{kri:but17}.

In this paper, we introduce a Bayesian hierarchical multilayer ERGM approach for ordinal/polytomous network data in order to simplify model specification and provide a substantial improvement in terms of interpretability by making use of binary ERGMs to capture the dependence structure between ordinal categories (network layers) of the weighted network structure.
The paper is organised as follows. In Section~\ref{sec:ergms}, we review the main features of ERGMs. In Section~\ref{sec:multilayer_rgm} we show how multilayer graphs can be used to represent weighted network structures. In Section~\ref{sec:multilayer_ergms} we introduce the multilayer ERGM approach. In Section~\ref{sec:relax_mergm} we focus on the interpretation of the multilayer ERGM framework as a dissolution process which is able to capture the between-layer generative process between the network layers. In Section~\ref{BayesianhmERGM} we generalise the modelling framework using a Bayesian hierarchical modelling approach and we propose to extend the approximate exchange algorithm \citep{cai:fri11} to sample from the doubly-intractable parameter posterior distribution (see \cite{par:hark18} for a recent review).
In Section~\ref{sec:sims}, we test our methodology on simulated data and, in Section~\ref{sec:apps} we demonstrate the usefulness of the multilayer ERGM approach by analysing two very well-known network datasets and assessing model fit by comparing goodness of fit diagnostics based on the weighted degree distribution. Final remarks are provided in the Conclusions in Section~\ref{sec:concl}.

\section{Exponential random graph models (ERGMs)}\label{sec:ergms}

Typically networks consist of a set of actors and relationships between pairs of them, for example, social interactions between individuals. The relational structure of a network graph is described by a random adjacency matrix $\bfY$ of a graph on $N$ nodes (actors) and a set of edges (relationships) $\{Y_{i,j}: i=1,\dots,N; j=1,\dots,N\}$ where $Y_{i,j} = 1$ if nodes $i$ and $j$ are connected and $Y_{i,j} = 0$ otherwise. The network graph $\bfY$ may be directed or undirected ($Y_{i,j} = Y_{j,i}$) depending on the nature of the relationships between the actors and $\bfy$ a realisation of $\bfY$. Self-loops are generally not allowed: $Y_{i,i} = 0.$

Exponential random graph models (ERGMs) are a particular class of discrete linear exponential families which represents the probability distribution of $\bfY$ as:

\begin{equation}
p(\bfy|\bftheta) = \frac{\exp\{\bftheta^\top s(\bfy)\}}{c(\bftheta)},
\label{eq:ergm}
\end{equation}

where $s(\bfy) \in \mathbb{R}^{\nstats}$ is a known vector of $\nstats$ sufficient statistics, $\bftheta \in \mathbb{R}^{\nstats}$ is the associated parameter vector, and $c(\bftheta)$ a normalising constant which is difficult to evaluate for all but trivially small graphs. The dependence hypothesis at the basis of these models is that edges self organise into small structures called configurations. There is a wide range of possible network configurations \citep{rob:pat:kal:lus07} which gives flexibility to adapt to different contexts. A positive parameter value for $\theta_i$ results in a tendency for the certain configuration corresponding to $s_i(\bfy)$ to be observed in the data more frequently than would otherwise be expected by chance.

\subsection{Curved ERGMs}\label{sec:cergms}

The ERGM likelihood defined in Equation~\ref{eq:ergm} can be generalised by assuming that the parameter $\bfxi(\bftheta)$   is nonlinear in the exponential family of distribution 
\begin{equation*}
p(\bfy | \bftheta) = \exp \left\{ \bfxi(\bftheta)^\top s(\bfy) - c \left( \bfxi(\bftheta)\right)\right\},
\label{eq:cergm}
\end{equation*}
which is therefore a curved exponential family \citep{hun07}. 
Curved ERGMs are implemented by specifying geometrically weighted network statistics \citep{sni:pat:rob:han06} and are commonly used to alleviate ERGM degeneracy issues \citep{han03}. However estimating the decay parameters associated to these network effects is a challenging issue and therefore these parameters are generally fixed rather than estimated. Recent contributions by \cite{bom:ban:hun16} and \cite{ste:sch:boj:mor19} have shown how curved ERGMs can be estimated in the context of Bayesian inference of bipartite and classical inference for multilevel network data, respectively.

\section{Weighted networks as multilayer graphs}\label{sec:multilayer_rgm}

A weighted network can be described by an $N \times N$ weighted adjacency matrix $\bfY$ where $Y_{i,j} \ne 0$ if nodes $i$ and $j$ are connected and $Y_{i,j} = 0$ otherwise. The value of $Y_{i,j}$ can be positive or negative, ordinal, count, bounded continuous and unbounded continuous. We denote with $D$ and $E$ the number of dyads and (weighted) edges of the network respectively.

\begin{figure}[htp]
\centering
\includegraphics[scale=0.5]{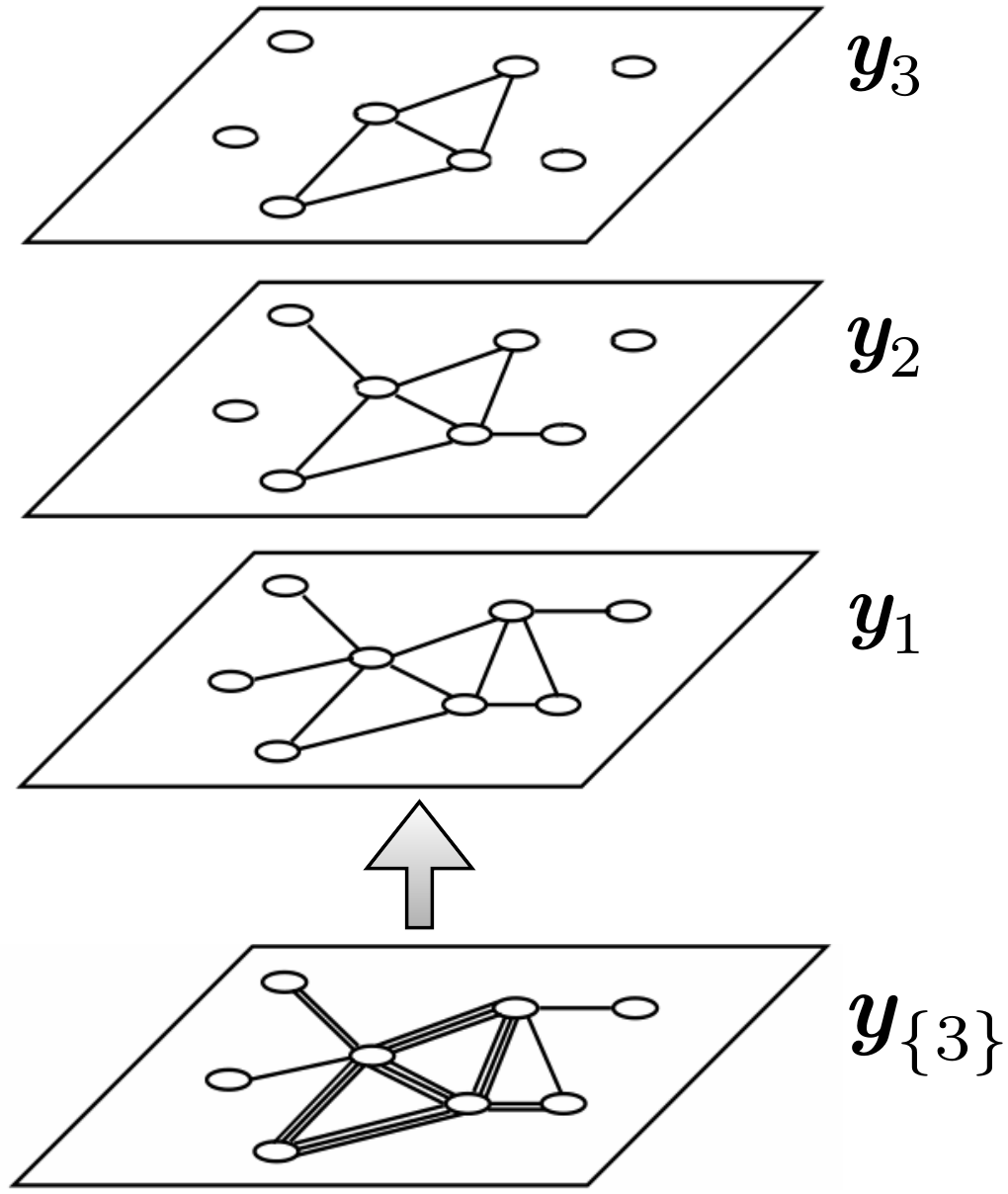}
\caption{An example of multilayer structure of a weighted network. The network $\bfy_{\{3\}}$ consists of edges that can take three ordinal values and its connectivity structure can be described by three overlapping binary network layers.}
\label{fig:net_layers}
\end{figure}

In this paper, we focus on ordinal and polytomous networks and we describe their connectivity structure through three-dimensional adjacency arrays \citep{robins1999logit} corresponding to what we define as a {\it multilayer network structure.} Let $\bfY$ be a weighted network on $N$ nodes with edges that are ordinal random variables taking value $w = 1, 2, \dots, W.$ We consider a three-dimensional array $N \times N \times W$ that we denote with $\bfy_{\{W\}},$ consisting of a set of {\it network layers} $\{ \bfY_{w},\; w = 1, \dots, W\}$ each of which represents a binary adjacency matrix with dyads defined as follows:
\begin{equation*}
Y_{i,j,w} 
  =\left\{\begin{matrix}
          1, & y_{i,j} \ge w; \\
          0, & y_{i,j} < w.\\
          \end{matrix}
   \right.
\end{equation*}
The set of edges observed in the $w$-th layer, $\mathcal{E}_w$, is a subset of the edges observed in the lower network layers, i.e.: $\mathcal{D}_1 \supset \mathcal{E}_1 \supset \mathcal{E}_2 \supset \dots \supset \mathcal{E}_W,$ where $\mathcal{D}_1$ is the set of dyads of the weighted network~(Figure \ref{fig:net_layers}).

\section{Multilayer ERGMs}\label{sec:multilayer_ergms}

The basic assumption of the multilayer network model is that the probability of observing the overall multilayer relational structure $\bfy_{\{W\}}$ corresponds to the product of conditional binary network models describing the transition processes between consecutive network layers $\bfy_{w}$ and $\bfy_{w+1}$: 

\begin{equation}\label{eq:multilayer}
p(\bfy_{\{W\}}| \bftheta) = p(\bfy_{1}, \bfy_{2}, \dots, \bfy_{W} | \bftheta)
= p(\bfy_{1} | \bftheta)\prod_{w = 1}^{W}p(\bfy_{w+1} | \bfy_{w}, \bftheta),
\end{equation}

where $\bftheta$ is a vector of parameters, $p(\bfy_{1} | \bftheta)$ is the unconditional network model for the first layer $\bfy_1$, $p(\bfy_{w+1} | \bfy_{w}, \bftheta)$ is the transition probability from layer $\bfy_w$ to layer $\bfy_{w+1}$ by assuming first order Markov dependence between layers (this assumption may be generalised to higher order Markov dependence), and $p(\bfy_{W+1} | \bfy_{W}, \bftheta)$ is the transition probability from the last layer $\bfy_W$ to the completely empty network graph $\bfy_{W + 1}$ (i.e., $y_{i,j,W + 1} = 0, \forall i,j \in \mathcal{D}_1$).

\subsection{The multilayer random graph model}\label{sec:mRGM}

The simplest multilayer network model is the multilayer random graph model which is based on the assumption of dyadic independence. Let $p$ denote the probability that two nodes $i$ and $j$ are connected in a network layer $w$ and $q = 1 - p.$ The multilayer random graph model can be defined as:
\begin{equation}
\begin{cases}\label{eq:RMGM}
p(\bfy_{1} | p) = p^{E_1} q^{D_1 - E_1}; \\
p(\bfy_{w+1} | \bfy_{w}, p) = p^{E_{w+1}}q^{E_{w} - E_{w+1}} = p^{E_{w+1}} q^{D_{w+1} - E_{w+1}}, \quad w = 1, \dots, W; 
\end{cases}
\end{equation}
where $D_1$ is the overall number of dyads in the weighted network and $E_{w}$ is the number of edges in the network layer $\bfy_w$ which corresponds to the number of random dyads $D_{w+1}$ allowed to take value 1 in the network layer $\bfy_{w+1}.$

The multilayer random graph model defined in Equation~\ref{eq:RMGM} corresponds to the weighted random graph model defined by \cite{gar09} which is equivalent to the geometric-reference random graph model defined by \cite{kri12}. In fact, merging Equations \ref{eq:multilayer} and \ref{eq:RMGM}, we have that:

$$p(\bfy_{\{W\}}| p) 
	= \prod_{w = 1}^{W} p^{E_w} q^{D_w - E_w} q^{E_W}
	= p^{\sum_{w = 1}^{W}E_w} 
	  q^{\sum_{w = 1}^{W}D_w - \sum_{w = 1}^{W - 1} E_w}.$$
	
Since: 
\begin{equation*}
\sum_{w = 1}^{W}D_w = D_1 + \sum_{w = 2}^{W}D_w = D_1 + \sum_{w = 1}^{W-1}E_w,
\end{equation*}
we have that:

$$
p(\bfy_{\{W\}}| p) 
	= p^{\sum_{w = 1}^{W}E_w} q^{D_1}
    = p^{\sum_{w = 1}^{W}\sum_{i,j \in \mathcal{D}_1}y_{i,j,w}} q^{D_1}
	= \prod_{(i,j) \in \mathcal{D}_1}p^{\sum_{w = 1}^{W}y_{i,j,w}} q,
$$

which corresponds to the product of the probabilities that any two nodes $(i,j)$ are connected by an edge of weight $\sum_{w = 1}^{W}y_{i,j,w} = y_{i,j}$ and therefore each weighted dyad $Y_{i,j} \overset{i.i.d.}{\sim} Geometric(q).$ 

The exponential form of the geometric-reference random graph model defined in Equation~\ref{eq:RMGM} is:
\begin{align}
p(\bfy_{\{W\}}| \theta) 
	&= \prod_{(i,j) \in \mathcal{D}_1} \frac{\exp\left\{ \theta\ y_{i,j} \right\}}{1 - \exp(\theta)}
     = \dfrac{\exp\left\{ \theta\ \sum_{(i,j) \in \mathcal{D}_1}y_{i,j} \right\}}{c(\theta)}, \notag
\end{align}

where $c(\theta)$ is a normalising constant, $1 - \exp(\theta) = q$ and $\exp(\theta) = p$ and therefore $\theta = \ln(p) \leq 0.$ 

\subsection{Multilayer ERGMs as dissolution processes}\label{sec:mERGM_diss}

The multilayer random graph model defined in Equation~\ref{eq:RMGM} can be easily extended by adopting a general multilayer ERGM generative process incorporating extra-dyadic network effects (such as degree-based statistics or transitive configurations) and thus relaxing the simplistic dyadic dependence assumption of the multilayer random graph model.
In Table~\ref{tab:transitions}, we can observe that edges between two nodes in layer $\bfy_w$ may or may not `survive' to the next upper layer $\bfy_{w+1}$ whereas empty edges in layer $\bfy_w$ remain empty edges in all the upper layers.

\begin{table}[h]
\centering
\caption{Possible transitions of a single dyadic variable between two consecutive network layers.}
\label{tab:transitions}
\begin{tabular}{ccc}
$Y_{i,j,w}$ & $\rightarrow$ & $Y_{i,j,w+1}$ \\ \hline
0           & $\rightarrow$ & 0             \\
1           & $\rightarrow$ & 0             \\
1           & $\rightarrow$ & 1             \\ \hline        
\end{tabular}
\end{table}

We assume that the transition process between consecutive network layers is a conditional ERGM model
$$
p(\bfy_{w+1}| \bfy_{w}, \bfphi) = \dfrac{ \exp \{ \bfphi^\top s(\bfy_{w+1};\bfy_{w}) \} }{c(\bfphi, \bfy_{w})},
$$
where $\bfphi$ is a parameter vector associated to $s(\bfy_{w+1};\bfy_{w})$ that is the vector of network statistics that do not dissolve in the transition process from $\bfy_{w}$ to $\bfy_{w+1}.$ This vector consists of standard binary network statistics such as the number of edges, stars, triangles, etc.\ that are present in both network layers $\bfy_{w}$ and $\bfy_{w+1}.$

The transition process from the last network layer $\bfy_{W}$ to the empty network graph $\bfy_{W + 1}$ is deterministic as $Y_{i,j,W + 1} = 0, \forall i,j \in \mathcal{D}_1$ regardless the value of $Y_{i,j,W}$ so it is not included in the model.

The conditional log-odds of an edge between nodes $i$ and $j$ in layer $\bfy_{w+1}$, given the presence of an edge between $i$ and $j$ in layer $\bfy_{w}$ while keeping all the rest of the network layer $\bfy_{w + 1}$ fixed is:
\begin{equation}\label{eq:log-odds1}
\ln \left( \frac{\Pr(Y_{i,j,w+1} = 1\ |\ Y_{i,j,w} = 1, \bfY_{-(i,j,w+1)})}
                 {\Pr(Y_{i,j,w+1} = 0\ |\ Y_{i,j,w} = 1, \bfY_{-(i,j,w+1)})} 
     \right)
     =\bfphi^\top \Delta(\bfy)_{i,j,w+1},
\end{equation}
where $\bfphi$ is a vector of parameters and $\Delta(\bfy)_{i,j,w+1}$ is the vector of change network statistics from $\bfy_{w}$ to $\bfy_{w + 1},$ i.e., the change in the value of the network statistic $s(\cdot)$ that would occur if $y_{i,j,w+1}$ were changed from 0 to 1 while leaving all of the rest of $\bfy_{w + 1}$ fixed.

From Table~\ref{tab:transitions}, we can notice that the formation process is not allowed when moving from layer $\bfy_{w}$ to layer $\bfy_{w + 1},$ so the conditional log-odds of an edge between nodes $i$ and $j$ in layer $\bfy_{w + 1},$ given the absence of an edge between nodes $i$ and $j$ in layer $\bfy_{w}$ is:
\begin{equation}\label{eq:log-odds2}
\ln \left( \frac{\Pr(Y_{i,j,w+1} = 1\ |\ Y_{i,j,w} = 0, \bfY_{-(i,j,w+1)})}
                 {\Pr(Y_{i,j,w+1} = 0\ |\ Y_{i,j,w} = 0, \bfY_{-(i,j,w+1)})} 
     \right)
     = -\infty,
\end{equation}
as $\Pr\left(Y_{i,j,w+1} = 1\ |\ Y_{i,j,w} = 0, \bfY_{-(i,j,w+1)}\right) = 0$ and $\Pr\left(Y_{i,j,w+1} = 0\ |\ Y_{i,j,w} = 0, \bfY_{-(i,j,w+1)}\right) = 1$. Therefore the parameter $\bfphi$ associated with the network effects expressed by the network statistics $s(\cdot)$ provides insights about the contribution of each network statistic to edge dissolution between consecutive network layers.

If we consider the multilayer random graph model defined in Section~\ref{sec:mRGM}, where only the number of edges is included in the model, we obtain $\Delta(\bfy)_{i,j,w+1} = 1$ and therefore the relationship between $\phi$ and $\theta$ is: $\phi = \ln(p) - \ln(q) = \theta - \ln(1 - \exp(\theta)).$

The likelihood of multilayer ERGMs defined by Equations~\ref{eq:log-odds1} and~\ref{eq:log-odds2} can be written as:
$$
p(\bfy_{\{W\}}| \bfphi) = \dfrac{ \exp \{ \bfphi^\top s(\bfy_{\{W\}}) \} }{c(\bfphi, \bfy_{\{W\}})}
= \dfrac{ \exp \{ \bfphi^\top s(\bfy_1) \} }{c(\bfphi, \bfy_1)} \prod_{w = 1}^{W-1} \frac{ \exp \{ \bfphi^\top s(\bfy_{w+1};\bfy_{w}) \} }{c(\bfphi, \bfy_w) }.
$$

The interpretation of the multilayer ERGM as a dissolution process implies that many network statistics developed for ERGMs can be readily used within this modelling framework, retaining much of their interpretation. A positive value for the parameter $\bfphi_i$ corresponding to a particular network statistic $s_i(\cdot)$ increases the probability of observing that network statistic in the next upper layer or vice versa. This means that the dissolution process between two consecutive layers can be interpreted exactly like a standard binary ERGM process defined in Section~\ref{sec:ergms}.
An analogous representation of multilayer ERGMs consists in considering a formation process from layer $W$ to 1 by just reversing the direction of the dependence structure between layers.

The multilayer ERGMs is similar to a static version of discrete separable temporal ERGMs \citep{kri14} which is including  initial conditions (i.e., the baseline network connectivity structure represented by the first network layer $\bfy_1$, see also the longitudinal ERGM approach developed by \cite{kos:cai:lom15}) but without formation process (which is not needed in this weighted network context).

\subsection{Relaxing the homogeneity assumption across layers}\label{sec:relax_mergm}

We now generalise the multilayer model introduced above by relaxing the parameter homogeneity assumption across network layers by considering layer-specific ERGM processes:
\begin{equation*}
p(\bfy_{\{W\}} | \bfphi_{1}, \dots, \bfphi_{W})
= p(\bfy_{1} | \bfphi_1)\prod_{w = 1}^{W-1}p(\bfy_{w+1} | \bfy_{w}, \bfphi_{w+1}).
\end{equation*}
where $\bfphi_{1}, \dots, \bfphi_{W}$ are the between-layer parameters capturing specific network effects that might characterise the transition processes between consecutive network layers. In fact, the behaviour of some network effects might vary depending on the layer values: we can for example immagine that for lower layers some network effects are positive and for higher layers the same effects are negative, or vice versa.

It is important to notice that the first ERGM transition $p(\bfy_{1} | \bfphi_1)$ is a standard binary ERGM that is equivalent to an ERGM dissolution process defined in Section~\ref{sec:mERGM_diss} conditional on the fully connected binary network graph that we denote by $\bfy_{0}$ defined on the same set of nodes, i.e., $y_{i,j,0} = 1, \forall i,j \in \mathcal{D}_1.$ Consequently, we have that $p(\bfy_{1} | \bfphi_1) = p(\bfy_{1} | \bfy_{0}, \bfphi_1),$ as: 
\begin{equation*}
\ln \left( \frac{\Pr(Y_{i,j} \ge 1\ |\ \bfY_{-(i,j)})}
                 {\Pr(Y_{i,j} = 0\ |\ \bfY_{-(i,j)})} 
     \right)
     =
\ln \left( \frac{\Pr(Y_{i,j,1} = 1\ |\ y_{i,j,0} = 1, \bfY_{-(i,j,1)})}
                 {\Pr(Y_{i,j,1} = 0\ |\ y_{i,j,0} = 1, \bfY_{-(i,j,1)})} 
     \right).
\end{equation*}

This modelling approach yields a flexible way to specify between-layer transition processes. In fact, it is possible to specify different network statistics for modelling different network transitions.

The conditional log-odds of an edge with weight $w^*$ between nodes $i$ and $j$ is:
\begin{equation*}
\ln \left( \frac{\Pr(Y_{i,j} = w^*\ |\ \bfY_{-(i,j)})}
                 {\Pr(Y_{i,j} = 0\ |\ \bfY_{-(i,j)})} 
     \right)
     =\sum_{w = 1}^{w^*}\bfphi_w^\top \Delta(\bfy)_{i,j,w}\ .
\end{equation*}

\section{Bayesian inference}\label{BayesianhmERGM}

Bayesian analysis is a promising approach to social network analysis because it yields a rich fully-probabilistic evaluation of uncertainty which is essential when dealing with complex and heterogenous relational data. 

The growing popularity of Bayesian techniques for ERGMs can be attributed to the development of efficient computational methods \citep{cai:fri11,alq:fri:eve:bol14,cai:mir15,bou:fri:mai17,bou:fri:mai18} and the availability of user-friendly software tools such as the {\sf Bergm} \citep{cai:fri14,Bergm_R} and {\sf hergm} \citep{hergm,hergm_R} packages for {\sf R} \citep{R_R}.

Using a Bayesian framework leads directly to the inclusion of prior information about the network effects into the modelling framework, and provides immediate access to the uncertainties by evaluating the posterior distribution of the parameters associated with the network effects. In social network analysis the Bayesian approach leads to the possibility of specifying informative parameter prior distribution consistent with some a priori expectation, for example, in terms of low density and high transitivity \citep{cai:pal:lom17}. In fact, parameter prior distribution can be concentrated on negative values for the density parameter and  positive values for transitivity parameters and/or  positive correlation between density and transitivity parameters.

In the following sections we will be extending the modelling framework introduced in Section~\ref{sec:relax_mergm} and describing a Bayesian parameter estimation procedure based on the approximate exchange algorithm \citep{mur:gha04,cai:fri11} which can sample from the ERGM posterior distribution which is doubly-intractable as both ERGM likelihood and marginal likelihood are not available.

\subsection{A hierarchical framework}\label{hmERGM}

The Geometric-reference ERGM approach \citep{gar09,kri12} is not capable of modelling every between-layer dependence process in the network and, on the other hand, the multilayer modelling approach defined in Section~\ref{sec:relax_mergm} is not capable of capturing overall trends of the network effects across the entire multilayer structure. For this reason, in order to model both the relational processes simultaneously and improve the goodness of fit of our model, we specify a hierarchical multilayer ERGM where layer specific parameters $\bfphi_1,\dots, \bfphi_W$ are coupled through a random variable $\bfeta$ representing the overall across-layer trend of the $\nstats$ network effects of interest with prior distribution $p(\bfeta|\bfgamma_0)$ defined by the hyper-parameters $\bfgamma_0$. 

The benefit of using Bayesian hierarchical approaches for ERGMs has been shown in several contexts. In particular, \cite{sla:koe16} demonstrated how these methods can be useful to describe the systematic patterns within groups and how these structural patterns differ across groups in multilevel networks.

\begin{figure}[htp]
\centering
\includegraphics[scale=0.4]{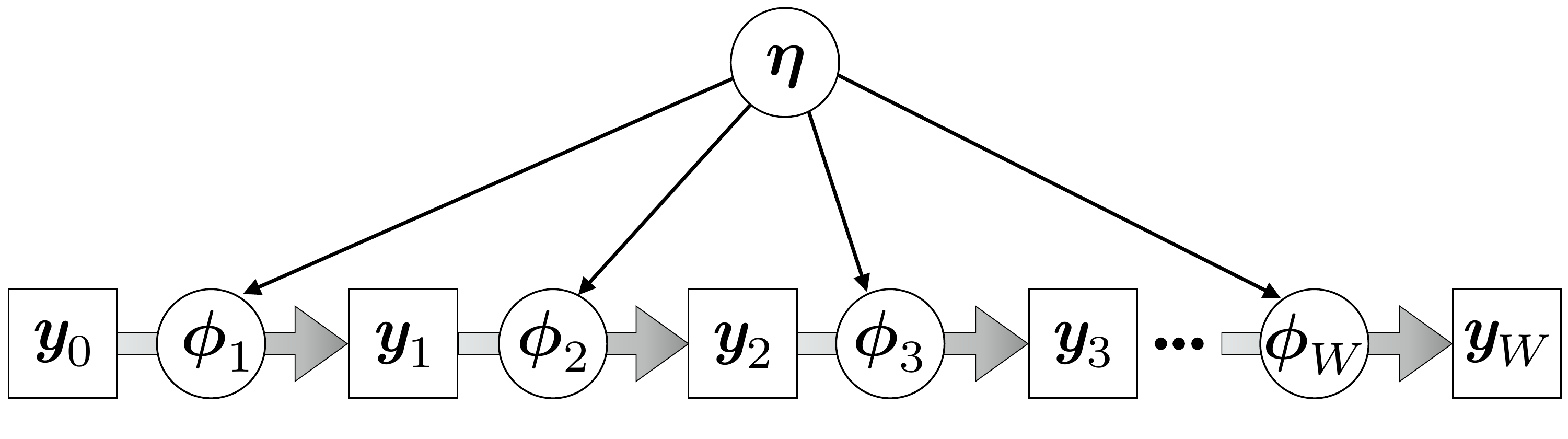}
\caption{Graphical structure of the hierarchical multilayer ERGM defined in Equation~\ref{eq:post_hmergm}.}
\label{fig:mERGM_framework}
\end{figure}

To develop a hierarchical multilayer approach, we define the following model:
\begin{equation}
p(\bfphi_{1}, \dots, \bfphi_{W}, \bfeta | \bfy_{\{W\}})
\propto 
  p(\bfy_{\{W\}} | \bfphi_{1}, \dots, \bfphi_{W})\
  p(\bfphi_1, \dots, \bfphi_W | \bfeta)\ p(\bfeta | \bfgamma_0)
  \label{eq:post_hmergm}
\end{equation}
and its structure is displayed in Figure~\ref{fig:mERGM_framework}.

We define a $\nstats$-dimensional model assuming that the between layer parameters are independent realisations from a Normal distribution, i.e.: 
$$p(\bfphi_1, \dots, \bfphi_W | \bfeta) = \prod_{w=1}^{W} p(\bfphi_w | \bfmu, \bfSigma),$$

where: $\bfphi_w | \bfmu, \bfSigma \sim \mathcal{N}(\bfmu, \bfSigma).$

We assume a Normal-Inverse-Wishart prior setting where: 
$$\bfSigma \sim \mathcal{W}^{-1}(\bfLambda_0, \nu_0); \quad \bfmu|\bfSigma \sim \mathcal{N}\left(\bfmu_0, \frac{\bfSigma}{\kappa_0}\right),$$ 
so that the joint prior distribution for $\bfmu, \bfSigma$ is: 
$$\bfmu, \bfSigma \sim \mathcal{NW}^{-1}(\bfmu_0, \kappa_0, \bfLambda_0, \nu_0).$$ 

The full conditional distribution of $\bfmu, \bfSigma$ is:
$$\bfmu, \bfSigma | \bfphi_1, \dots, \bfphi_W \sim \mathcal{NW}^{-1}(\bfmu_1, \kappa_1, \bfLambda_1, \nu_1);$$ 
with parameters:
$$\bfmu_1 = \frac{\kappa_0}{\kappa_0 + W}\bfmu_0 + \frac{W}{\kappa_0 + W}\bar{\bfphi};\quad 
\kappa_1 = \kappa_0 + W;$$
$$\bfLambda_1 = \bfLambda_0 + \bfS + \frac{\kappa_0\ W}{\kappa_0 + W}(\bar{\bfphi} - \bfmu_0)(\bar{\bfphi} - \bfmu_0)^\top;\quad 
\nu_1 = \nu_0 + W;$$
where: 
$$\bar{\bfphi} = \frac{1}{W}\sum_{w=1}^{W} \bfphi_w;\quad  \bfS = \sum_{w=1}^{W}(\bfphi_w - \bar{\bfphi})(\bfphi_w - \bar{\bfphi})^\top.$$

\subsection{Posterior estimation}

To estimate the posterior distribution $p(\bfphi_{1}, \dots, \bfphi_{W}, \bfmu, \bfSigma | \bfy_{\{W\}})$ defined in Equation~\ref{eq:post_hmergm}, we extend the approximate exchange algorithm of \cite{cai:fri11} to sample from the doubly-intractable distribution $p(\bfphi_{1}, \dots, \bfphi_{W} | \bfy_{\{W\}}, \bfmu, \bfSigma)$ and then we sample $\bfmu$ and $\bfSigma$ from the full conditional distribution via Gibbs sampling.

The approximate exchange algorithm, described in Algorithm~\ref{alg:AEA}, is an asymptotically exact MCMC algorithm as it guarantees asymptotically exact recovery of the posterior distribution as the number of posterior samples increases \citep{eve12}. 

\begin{algorithm}
\caption{Approximate exchange algorithm for $p(\bfphi_{1}, \dots, \bfphi_{W} | \bfy_{\{W\}}, \bfmu, \bfSigma)$}
\begin{algorithmic}\label{alg:AEA}
\STATE {Initialise $\bfphi_{1}^{(1)}, \dots, \bfphi_{W}^{(1)}$}
\FOR{$i = 1, \dots, I$}
\FOR{$w = 1, \dots, W$} 
	\STATE {\bf 1)} {$\bfphi_w' \sim h(\cdot | \bfphi_w^{(i)})$} 
	\STATE {\bf 2)} {$\bfy'_{w} \sim p(\cdot| \bfphi_w', \bfy_{w-1})$ via MCMC (see Algorithm 2)} 
	\STATE {\bf 3)} {Set $\bfphi_w^{(i+1)} = \bfphi_w'$ with probability:}
	\STATE          {$$\min \left( 1, 
	                             \dfrac{ 
	                                   q_{\bfphi_w^{(i)}}(\bfy'_{w}; \bfy_{w-1})\ 
	                                   p(\bfphi_{1}, \dots, \bfphi_w', \dots, \bfphi_{W}|\bfmu, \bfSigma)\ 
	                                   q_{\bfphi_w'}(\bfy_{w}; \bfy_{w-1})}
                                      {q_{\bfphi_w^{(i)}}(\bfy_{w}; \bfy_{w-1})\ 
                                       p(\bfphi_{1}, \dots, \bfphi_w^{(i)}, \dots, \bfphi_{W}|\bfmu, \bfSigma)\ 
                                       q_{\bfphi_w'}(\bfy'_{w}; \bfy_{w-1})}
                          \right)$$} 
\ENDFOR
\ENDFOR
\end{algorithmic}
\end{algorithm}

The unnormalised between-layer ERGM likelihood is defined as: 
$$
q_{\bfphi_w}(\bfy_w; \bfy_{w-1}) = \exp \{ \bfphi_w^\top s(\bfy_{w+1};\bfy_{w}) \},
$$
and $h(\cdot)$ is a proposal distribution for updating the model parameters. Adaptive strategies \citep{cai:fri11,cai:mir15} and approximate transition kernels approaches \citep{alq:fri:eve:bol14} have been successfully implemented in this context.

From a computational viewpoint, the algorithm becomes increasingly expensive as the network size and the number of layers increase. In fact, the number of MCMC iterations needed to simulate the auxiliary network layers at Step 2 of Algorithm~\ref{alg:AEA} is proportional to the number of dyads in the network \citep{eve12} and the MCMC sampling is repeated for generating the binary network layers of the observed adjacency array as many times as the value of maximum weight ($W$) that an edge can take.

However, it is important to notice that the number of iterations needed for simulating the network layer $\bfy'_{w+1}$ should be just proportional to the number of edges ($E_{w}$) in the lower network layer $\bfy_{w}$ which is a subset of the overall number of dyads ($\mathcal{D}_1$) in the network.
In fact, according to Equation~\ref{eq:log-odds1}, only the subset of dyads belonging to $\mathcal{D}_{w+1}$ and corresponding to the non-zero dyads of the network layer $\bfy_{w}$ can take value $w+1$ and they are the only dyads involved in the network simulation of layer $\bfy'_{w+1} \sim p(\cdot| \bfphi_{w+1}', \bfy_{w}).$ 

The constrained ERGM simulation for the network layer $\bfy_{w+1}$ corresponding to Step 2 of Algorithm~\ref{alg:AEA} is described in detail in Algorithm~\ref{alg:SIM}. The number of MCMC steps $I_{w+1}$ required for simulating a network layer $\bfy_{w+1}$ can be smaller than the number of steps required for simulating any lower layer and consequently the computational cost of Step 2 of Algorithm~\ref{alg:AEA} decreases for increasing values of $w.$

\begin{algorithm}
\caption{Constrained network layer simulation for $\bfy_{w+1}$}
\begin{algorithmic}\label{alg:SIM}
\STATE {Initialise $\bfy_{w+1}^{(1)} = \bfy_{w}$}
\FOR{$i = 1, \dots, I_{w+1}$}
	\STATE {\bf 1)} {$\bfy'_{w+1} \sim h(\cdot| \bfy_{w+1}^{(i)}, \bfy_{w})$} 
	\STATE {\bf 2)} {Set $\bfy^{(i+1)}_{w+1} = \bfy'_{w+1}$ with probability:}
	\STATE          {
	$$
	\begin{aligned}
	\min
    \left\lbrace 
    1,\dfrac{p(\bfy'_{w+1}\ |\ \bfphi_{w+1}, \bfy_{w})\ h(\bfy'_{w+1} | \bfy_{w+1}^{(i)}, \bfy_{w})}
            {p(\bfy^{(i)}_{w+1}\ |\ \bfphi_{w+1}, \bfy_{w})\ h(\bfy^{(i)}_{w+1} | \bfy'_{w+1}, \bfy_{w})}
    \right\rbrace 
    \end{aligned}$$} 
\ENDFOR
\end{algorithmic}
\end{algorithm}

In this paper, we adopt the `tie-no-tie' sampler for the proposal distribution $h(\cdot)$ which is the default procedure in the {\sf ergm} package \citep{hun:han:but:goo:mor08,ergm_R} for {\sf R} at Step 1 of Algorithm~\ref{alg:SIM}.

\section{Simulation study}\label{sec:sims}

We propose two simulation studies to assess the performance of the proposed modelling framework and the benefits of the hierarchical Bayesian approach. We consider two 50-node weighted networks with 3 network layers generated by two distinct plausible network processes. 
We specify the same set of ERGM network statistics for both datasets:
\begin{itemize}
\item[$s_1:$] Edge statistic ({\sf edges}) is the number of edges in the network and captures the network density effect.
\item[$s_2:$] Geometrically weighted edgewise shared partner statistic ({\sf gwesp}) is a function of the edgewise shared partner statistics $EP_d (\bfy)$ defined as the number of unordered connected pairs $(i, j)$ (partners) that are both connected to exactly $d$ other nodes: $\sum_d g_d (\alpha) \; EP_d (\bfy),$
where $g_d (\alpha)$ is an exponential weight function with decay parameter $\alpha$ defined as:
\begin{equation*}
g_d (\alpha) = e^{\alpha} \left\{ 1 - \left( 1 - e ^{-\alpha} \right)^{d} \right\}.
\end{equation*}
This statistic captures the tendency towards transitivity, i.e., the tendency of edges to be connected through multiple triadic relations simultaneously.
\end{itemize}

As mentioned in Section~\ref{sec:cergms} estimating the decay parameters $\alpha$ of the geometrically weighted network statistics from a single network is in general challenging. For this reason, we fix $\alpha = \log(2)$ for all the geometrically weighted network statistics used in this paper.

The between layer parameters $\bfphi$ of the two examples are respectively generated from the following prior distributions:
\begin{itemize}
\item[(a)] 
$\bfphi | \bfmu, \bfSigma \sim \mathcal{N}\left(
\bfmu = 
\begin{pmatrix}
0 \\ 
0
\end{pmatrix},
\bfSigma =  
\begin{pmatrix}
2 & 0 \\ 
0 & 1 
\end{pmatrix}
\right),$ meaning that both density ($\mu_1$) and transitivity ($\mu_2$) overall trends are null.
\item[(b)]
$\bfphi | \bfmu, \bfSigma \sim \mathcal{N}\left(
\bfmu = 
\begin{pmatrix}
-2 \\ 
0.5
\end{pmatrix}, 
\bfSigma =  
\begin{pmatrix}
2 & 0 \\ 
0 & 0.5 
\end{pmatrix}
\right),$ corresponding to negative density and positive transitivity overall trends.
\end{itemize}

\begin{figure}[htp]
\centering
\includegraphics[scale=0.55]{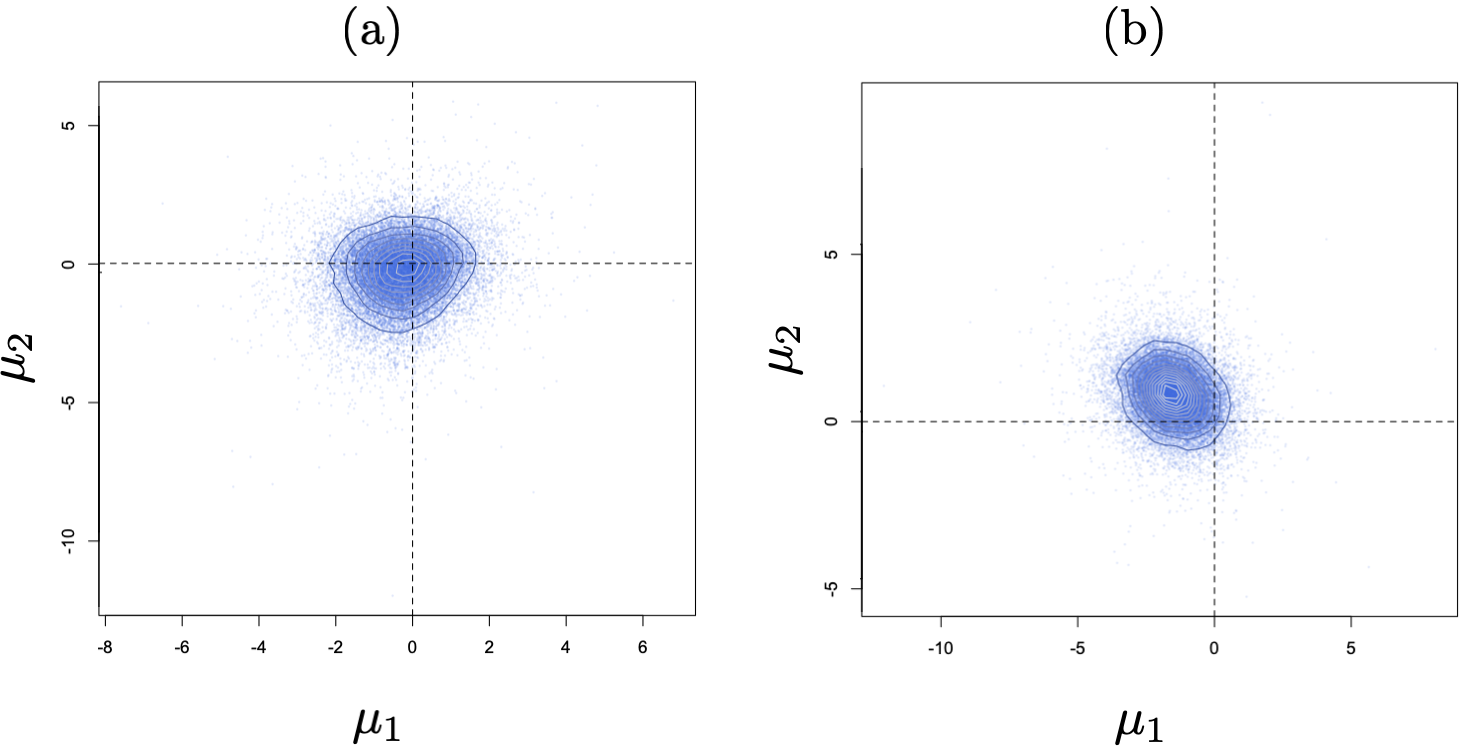}
\caption{Estimated posterior distribution for $\bfmu$ for each simulated network.}
\label{fig:sims}
\end{figure}

For both the simulated datasets we used the adaptive direction sampling strategy based on 4 MCMC chains consisting of 10,000 iterations each. We tuned the algorithm parameters in order to get about 20\% acceptance rate. In terms of hyper-parameters for the hierarchical model, we set $\kappa_0 = 1, \bfLambda_0 = I_{\nstats}, \nu_0 = \nstats + 2,$ where $I_{\nstats}$ is the $\nstats \times \nstats$ identity matrix ($r = 2$ in this case). An analogous hyper-prior specification will be used for the applications in Section~\ref{sec:apps}.

\begin{figure}[htp]
\centering
\includegraphics[scale=0.6]{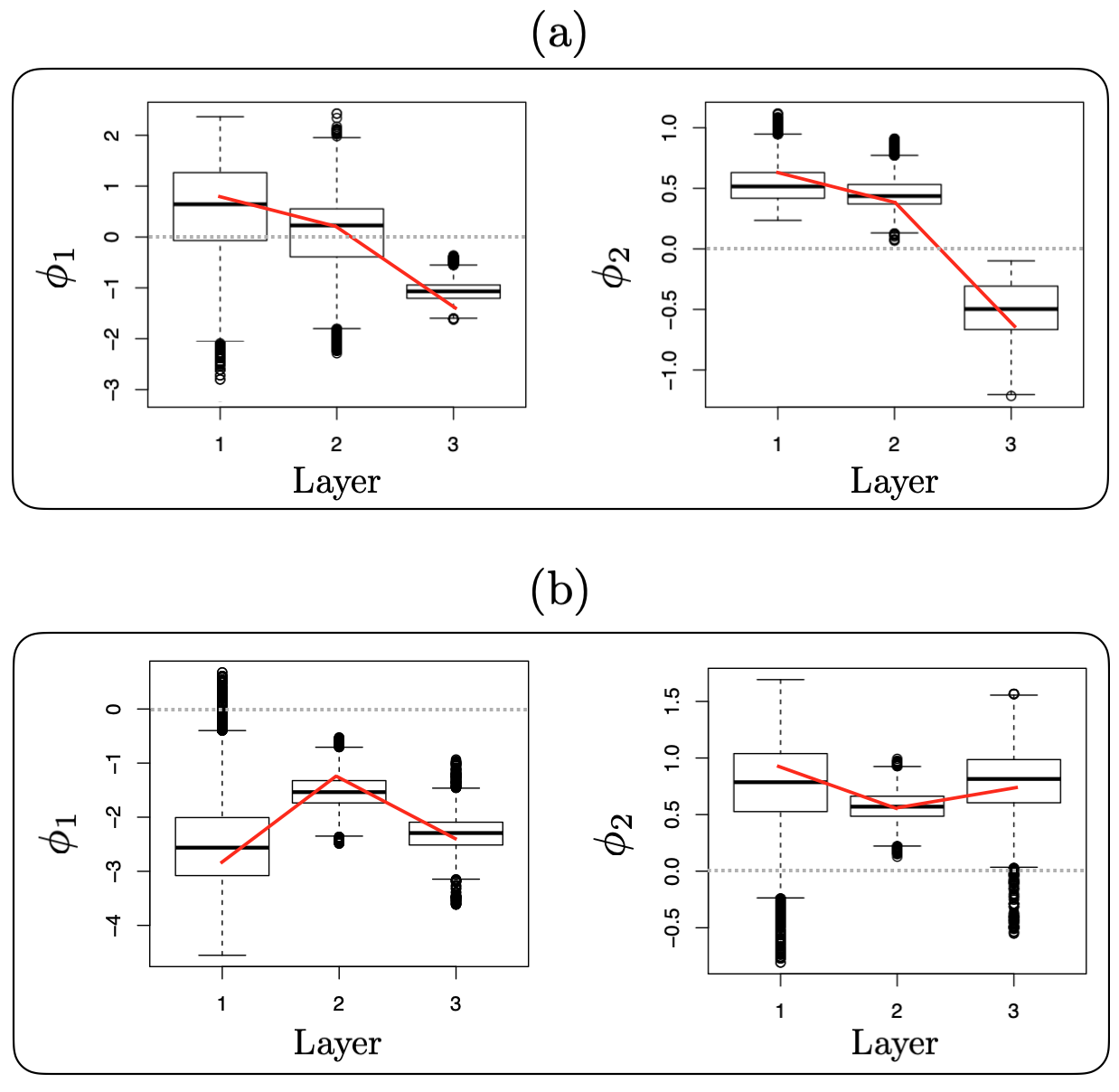}
\caption{The solid red line represents the simulated values of $\bfphi_1$ ({\sf edges}) and $\bfphi_2$  ({\sf gwesp}) for simulation study (a) and (b). The boxplots represent the estimated predictive posterior for $\bfphi$ for every network layer.}
\label{fig:sims_pp}
\end{figure}

As shown in Figure~\ref{fig:sims}, the estimated posterior modes correspond to the true values of $\bfmu$ for both examples.
The estimated predictive posterior densities (displayed in Figure~\ref{fig:sims_pp}) are in agreement with the simulated parameters $\bfphi$ for both the experiments. Moreover, it is possible to observe how the hierarchical multilayer model allows for the accounting of the structural between-layer variation and dependencies. In particular, in simulation study (a) even though the overall trends are null ($\mu_1 = \mu_2 = 0$) we can see that between-layer processes are heterogeneous and in the conditional process between network layer 2 and 3 both density and transitivity effects are negative in contrast to what happens in the lower between-layer processes where the transitivity effects are positive and the density effects are null.

\section{Applications}\label{sec:apps}

In this section we present two examples of application of the hierarchical multilayer modelling framework defined in Section~\ref{hmERGM}. We will be considering two well-known datasets: the Bernard and Killworth office network \citep{killworth} and the Zachary karate club network \citep{zac77}. We will focus on a polytomous transformation of weighted networks by thresholding the weighted edges at 3 different values so as to obtain 3-layer network structures where the strength of the dyadic relations can be interpreted as either low, medium, or high. This procedure has been carried out by arbitrarily defining threshold values, based on the quantiles of the edge weight distribution. 

\subsection{Bernard and Killworth office network}

The Bernard and Killworth office network dataset concerns the observed frequency of interactions between $N = 40$ individuals in a small business office.

\begin{figure}[htp]
\centering
\includegraphics[scale=0.37]{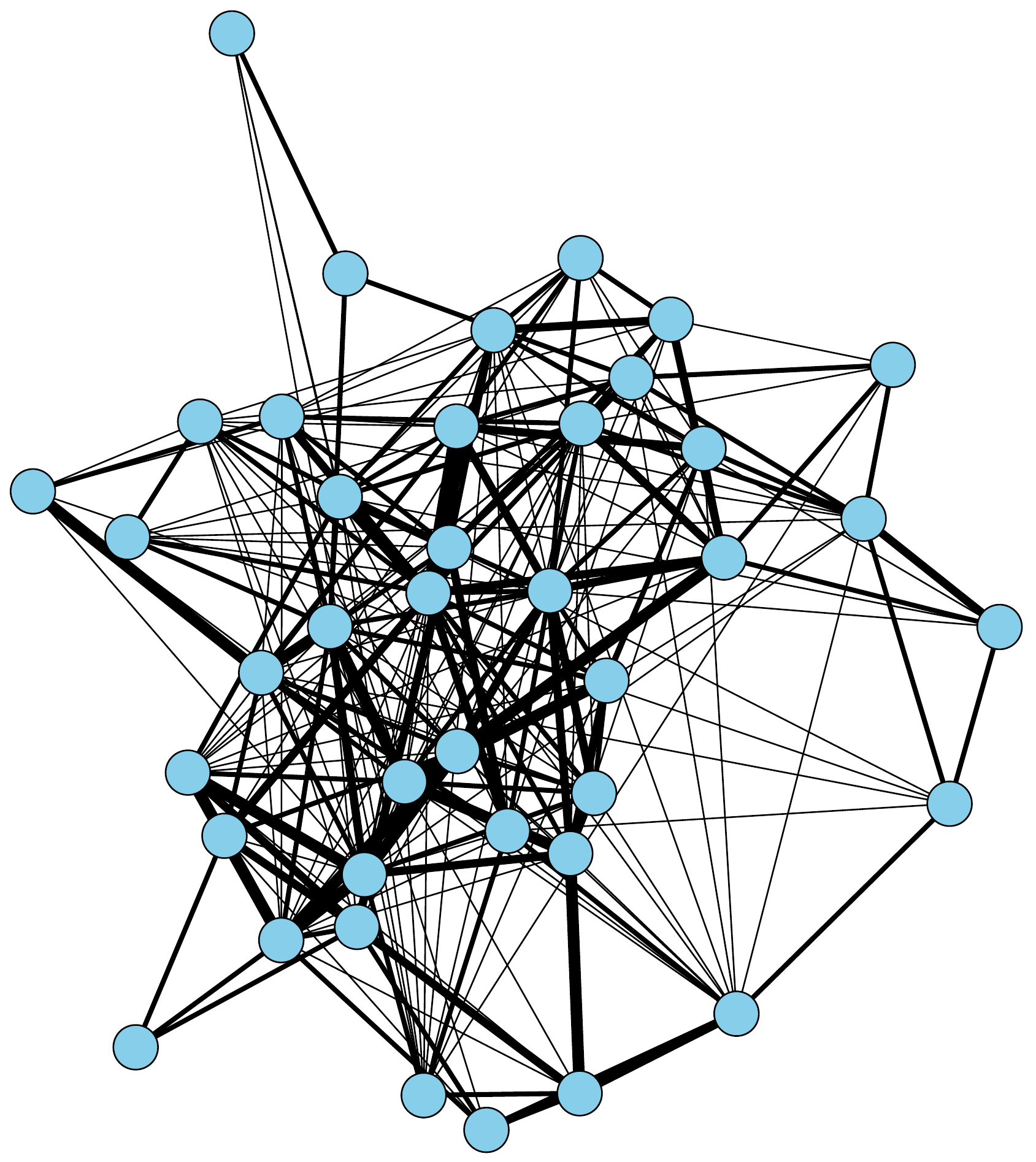}
\caption{Weighted structure of the Bernard and Killworth office network derived by thresholding the original weights at $2, 4,$ and  $8.$ The width of the edge lines is proportional to the edge weight.}
\label{fig:wbkoff}
\end{figure}

\begin{figure}[htp]
\centering
\includegraphics[scale=0.6]{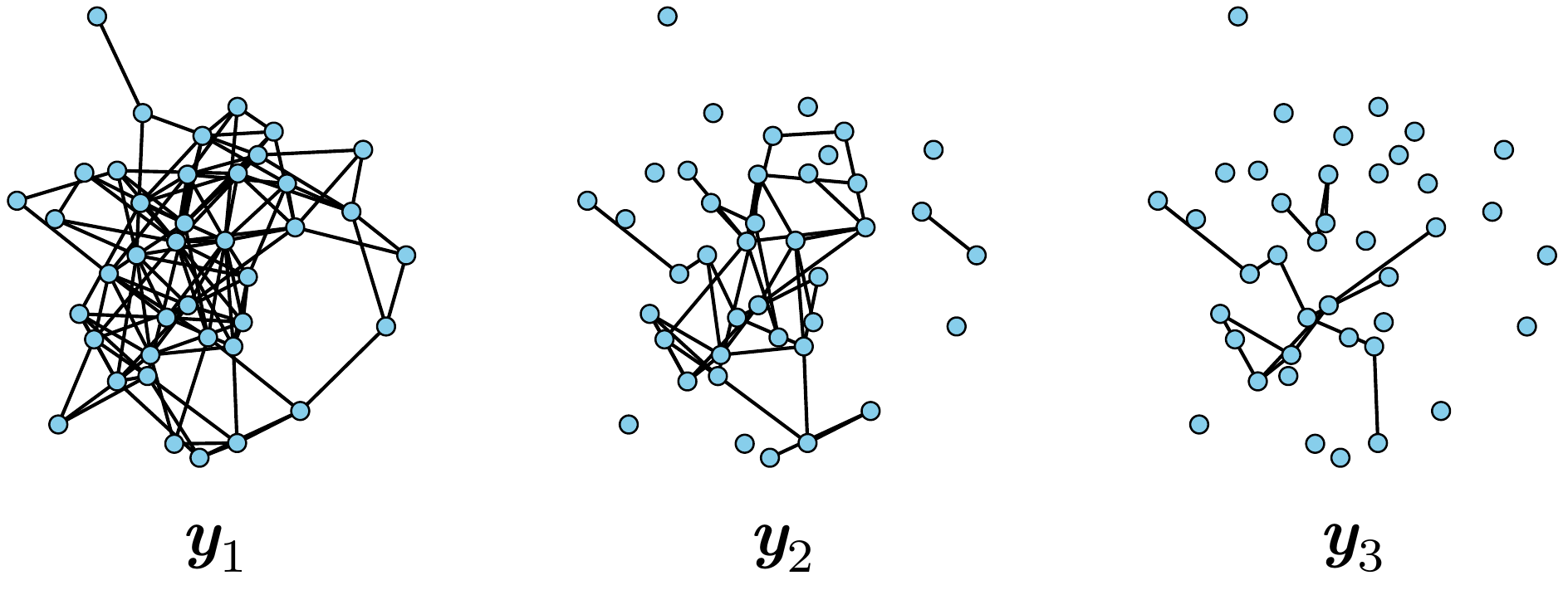}
\caption{The multilayer structure of the Bernard and Killworth office network. The number of edges in each network layer is: $E_1 = 123, E_2 = 44, E_3 = 15.$}
\label{fig:bkoff_mgraph}
\end{figure}

The network graph and the multilayer visualisation of the Bernard and Killworth office dataset obtained by selecting 3 layers of the weighted graph are shown in Figure~\ref{fig:wbkoff} and Figure~\ref{fig:bkoff_mgraph} respectively.

\subsubsection{Model specification}

As mentioned in Section~\ref{sec:mERGM_diss}, an important aspect of the multilayer modelling approach is that it can accomodate binary network statistics \citep{sni:pat:rob:han06} to describe the weighted network topology. For the Bernard and Killworth office dataset we propose the following model specification:
\begin{itemize}
\item[$s_1:$] Edge statistic ({\sf edges}).
\item[$s_2:$] Geometrically weighted degree statistic ({\sf gwdegree}):
$\sum_d g_d (\alpha) D_d (\bfy),$ where $g_d (\alpha)$ is the exponential weight function with decay parameter $\alpha.$ This statistic captures the tendency towards centralisation in the degree distribution $D_d (\bfy)$ of the network.
\item[$s_3:$] Geometrically weighted edgewise shared partner statistic ({\sf gwesp}).
\end{itemize}

\subsubsection{Posterior analysis}

We used the adaptive direction sampling strategy based on 6 MCMC chains consisting of 10,000 iterations each for improving the mixing of the approximate exchange. Traceplots for the model parameters can be found in the Appendix.

\begin{figure}[htp]
\centering
\includegraphics[scale=0.6]{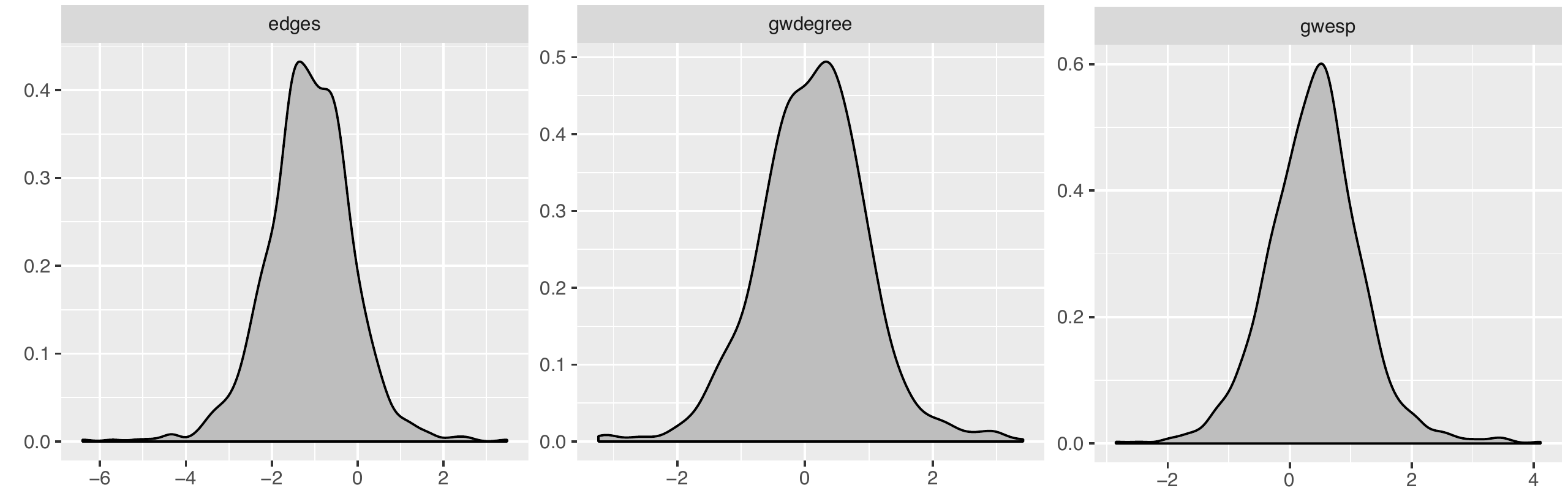}
\caption{Posterior distribution of $\mu_{1}$ ({\sf edges}), $\mu_{2}$ ({\sf gwdegree}), and $\mu_{3}$ ({\sf gwesp}) for the Bernard and Killworth office network.}
\label{fig:pMu_bkoff1}
\end{figure}

The posterior distribution for $\bfmu$, displayed in Figure~\ref{fig:pMu_bkoff1} and summarised in Table~\ref{tab:pMu_bkoff1}, shows the general tendencies of the three local effects across the network layers. A large part of the probability density of $\mu_1$ (corresponding to the {\sf edges} statistic) is concentrated on negative values. This is compensated by a large part of the probability density of $\mu_3$ (corresponding to the {\sf gwesp} statistic) mostly concentrated on positive values. These two tendencies explain the overall increasing sparsity and transitivity of the multilayer process across the network layers.

\begin{table}[htp]
\centering
\caption{Posterior estimates for $\bfmu$ for the Bernard and Killworth office network.}
\label{tab:pMu_bkoff1}
\begin{tabular}{lcc}
Parameter (Effect)       & Mean  & SD     \\ \hline
$\mu_1$ ({\sf edges})    & -1.10 & 1.00   \\
$\mu_2$ ({\sf gwdegree}) & 0.14  & 0.92   \\
$\mu_3$ ({\sf gwesp})    & 0.38  & 0.78   \\ \hline
\end{tabular}
\end{table}

\begin{figure}[htp]
\centering
\includegraphics[scale=0.5]{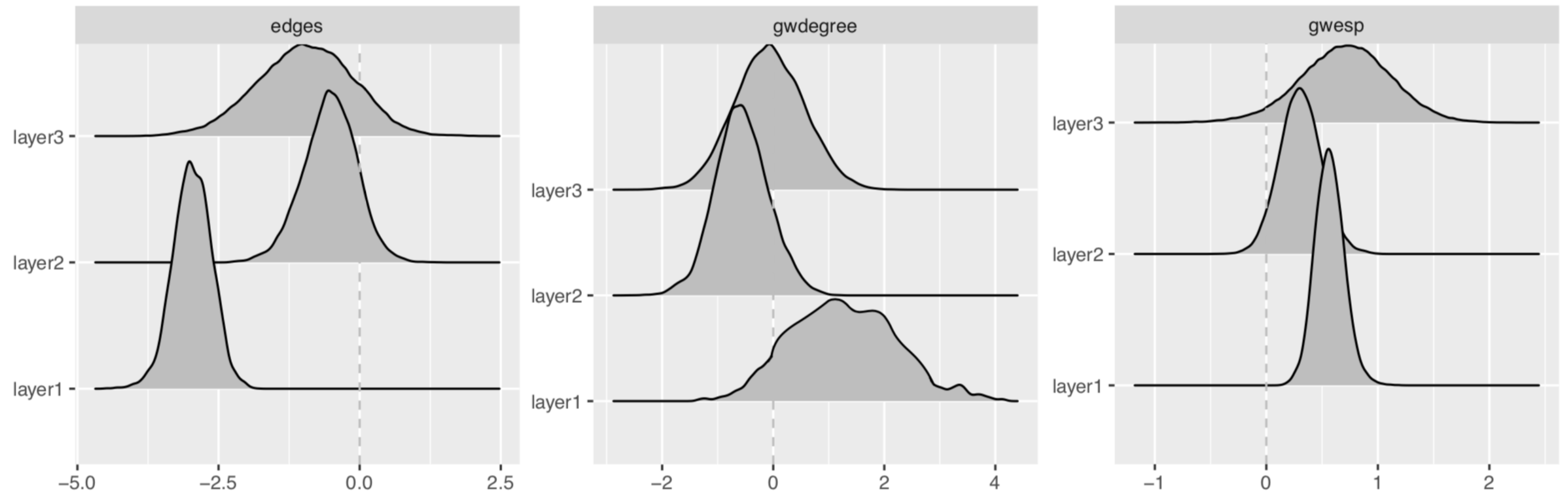}
\caption{Predictive posterior of $\bfphi_{1}$ ({\sf edges}), $\bfphi_{2}$ ({\sf gwdegree}), $\bfphi_{3}$ ({\sf gwesp}) for the Bernard and Killworth office network.}
\label{fig:ppPhi_bkoff1}
\end{figure}

The predictive posterior for $\bfphi_{1}, \bfphi_{2}, \bfphi_{3}$, displayed in Figure~\ref{fig:ppPhi_bkoff1} and summarised in Table~\ref{tab:ppPhi_bkoff1}, allows us to understand the between-layer ERGM processes. In particular we can notice that most of the density dissolution process described above is concerning the generation of the first network layer as the 99\% credible interval of $\bfphi_1$ (corresponding to the {\sf edges} statistic) falls completely on negative values. On the other hand the increase of transitivity captured by $\bfphi_3$ (corresponding to the {\sf gwesp} statistic) is observed between each layer meaning that the dissolution ERGM process across layers affects edges that are not embedded into transitive structures. The tendency towards centralisation in the degree distribution represented by $\mu_2$ and $\bfphi_2$ (corresponding to the {\sf gwdegree} statistic) does not seem to be important in explaining both the overall and the between-layer weighted structure of the network meaning that the strong edges are not necessarily centralised or dispersed in the degree distribution.

\begin{table}[htp]
\centering
\caption{Predictive posterior estimates of $\bfphi_{1}, \bfphi_{2}, \bfphi_{3}$ for the Bernard and Killworth office network.}
\label{tab:ppPhi_bkoff1}
\begin{tabular}{lcccccc}
                          & \multicolumn{2}{c}{Layer 1} & \multicolumn{2}{c}{Layer 2} & \multicolumn{2}{c}{Layer 3} \\
Parameter (Effect)        & Mean  & SD   & Mean  & SD   & Mean  & SD   \\ \hline
$\bfphi_1$ ({\sf edges})    & -2.94 & 0.34 & -0.53 & 0.47 & -0.94 & 0.85 \\
$\bfphi_2$ ({\sf gwdegree}) & 1.17  & 0.96 & -0.56 & 0.48 & -0.08 & 0.62 \\
$\bfphi_3$ ({\sf gwesp})    & 0.56  & 0.13 & 0.31  & 0.18 & 0.70  & 0.41 \\ \hline
\end{tabular}
\end{table}


\subsection{Zachary karate club network}

\begin{figure}[htp]
\centering
\includegraphics[scale=0.53]{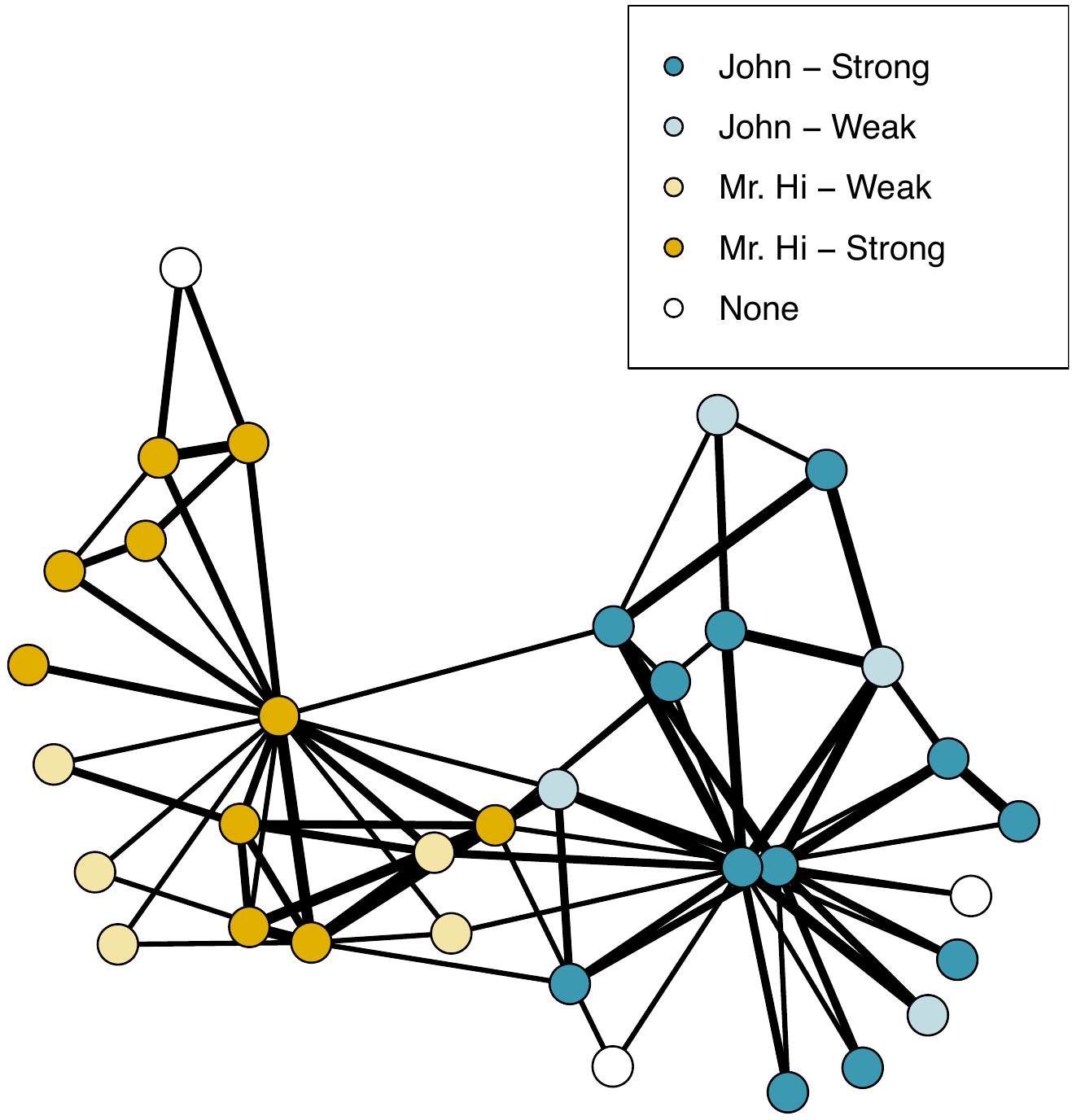}
\caption{Weighted graph structure of the Zachary karate club network derived by thresholding the original weights at $1, 3,$ and  $4.$ The width of the edge lines is proportional to the edge weight.}

\label{fig:wzach}
\end{figure}

The Zachary karate club network concerns social relations in a university karate club involving 34 individuals. The network graph in Figure~\ref{fig:wzach} shows the relative strength of the associations, i.e., the number of situations in and outside the club in which interactions occurred between individuals.

\begin{figure}[htp]
\centering
\includegraphics[scale=0.45]{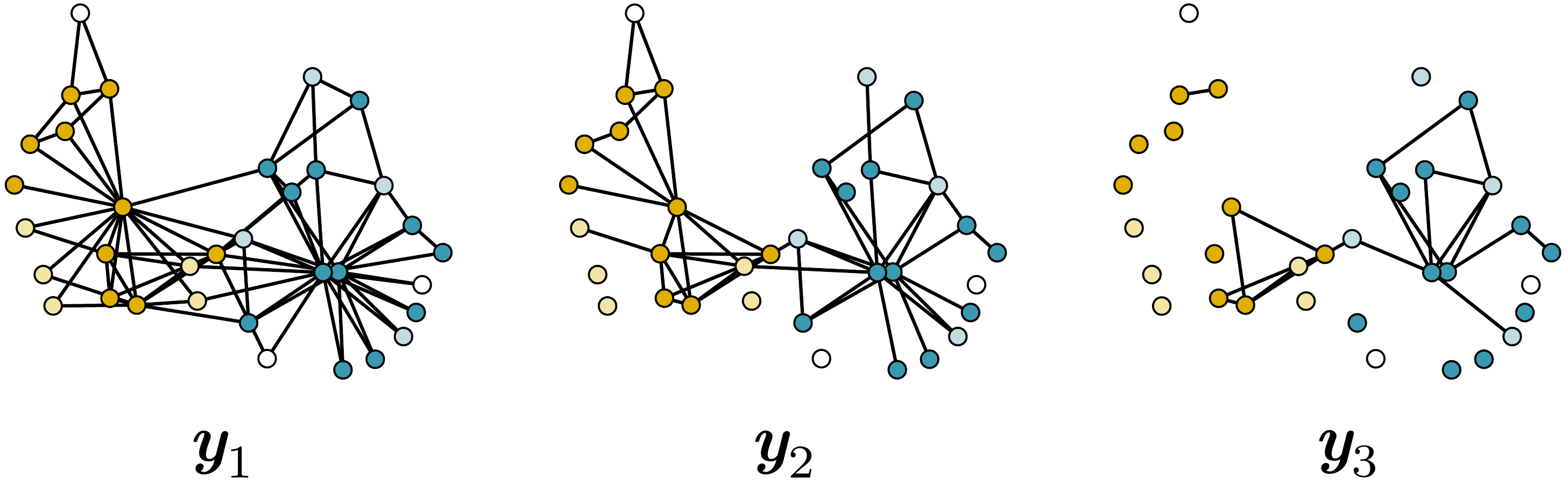}
\caption{The multilayer structure of the Zachary karate club weighted network. The number of edges in each layer is: $E_1 = 78, E_2 = 48, E_3 = 21.$}
\label{fig:zach_mgraph}
\end{figure}

A multilayer visualisation of the karate club network obtained by selecting 3 layers of the weighted graph is shown in Figure~\ref{fig:zach_mgraph}.

\subsubsection{Model specification}

We included in the model the {\sf edges} ($s_1$) and {\sf gwesp} ($s_2$) statistics as in the previous example plus the following statistics:
\begin{itemize}
\item[$s_3:$] Geometrically weighted non-edgewise shared partner statistic ({\sf gwnsp}) is a function of the non-edgewise shared partner statistics $NP_d (\bfy)$ defined as the number of unordered unconnected pairs $(i, j)$ to exactly $d$ other nodes: $\sum_d g_d (\alpha) \; NP_d (\bfy).$ This statistic captures the tendency of non-directly-connected nodes to be connected through multiple others.
\item[$s_4:$] Homophily statistic ({\sf nodematch}) is the number of edges between actors having the same nodal attribute $X.$ For this example, $X$ is the faction alignment of the club members so that this statistic captures the density of edges between nodes within the same faction.
\end{itemize}

\subsubsection{Posterior analysis}
We used the adaptive direction sampling strategy based on 8 MCMC chains consisting of 10,000 iterations each. In terms of hyper-parameters for the hierarchical model, we used the same set-up of the previous application.

\begin{figure}[htp]
\centering
\includegraphics[scale=0.6]{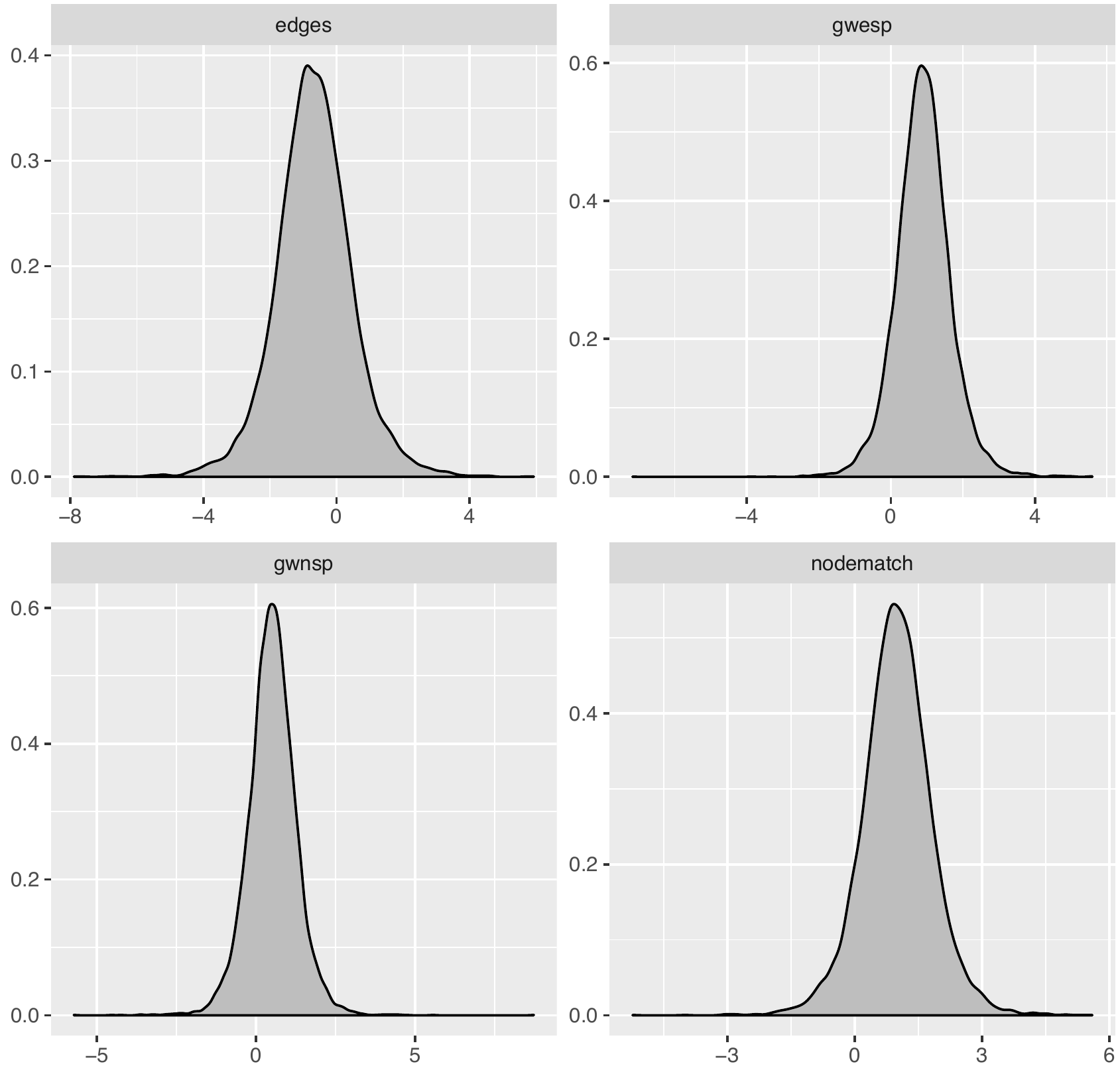}
\caption{Posterior distribution of $\mu_{1}$ ({\sf edges}), $\mu_{2}$ ({\sf gwesp}), $\mu_{3}$ ({\sf gwnsp}), and $\mu_{4}$ ({\sf nodematch}) for the Zachary karate club network.}
\label{fig:pMu_zach1}
\end{figure}

The posterior distribution for $\bfmu$, displayed in Figure~\ref{fig:pMu_zach1} and summarised in Table~\ref{tab:pMu_zach1}, shows the general tendencies of the 4 local effects across the 3 network layers. A large part of the probability density of $\mu_1$ (corresponding to the {\sf edges} statistic) is concentrated on negative values. This is compensated by a large part of the probability density of $\mu_2$ (corresponding to the {\sf gwesp} statistic) and $\mu_4$ (corresponding to the {\sf nodematch} statistic) mostly concentrated on positive values. The probability density of $\mu_3$ (corresponding to the {\sf gwnsp} statistic) is concentrated on values around 0.
These tendencies suggest that the overall dissolution process mainly concerns edges that do not connect nodes within factions and/or are not embedded in transitive triads.

\begin{table}[htp]
\centering
\caption{Posterior estimates of $\bfmu$ for the Zachary karate club network.}
\label{tab:pMu_zach1}
\begin{tabular}{lcc}
Parameter (Effect)        & Mean  & SD     \\ \hline
$\mu_1$ ({\sf edges})     & -1.20 & 1.16   \\
$\mu_2$ ({\sf gwesp})     & 0.40  & 0.77   \\
$\mu_3$ ({\sf gwnsp})     & 0.02  & 0.76   \\
$\mu_4$ ({\sf nodematch}) & 0.51  & 0.82   \\ \hline
\end{tabular}
\end{table}

\begin{figure}[htp]
\centering
\includegraphics[scale=0.6]{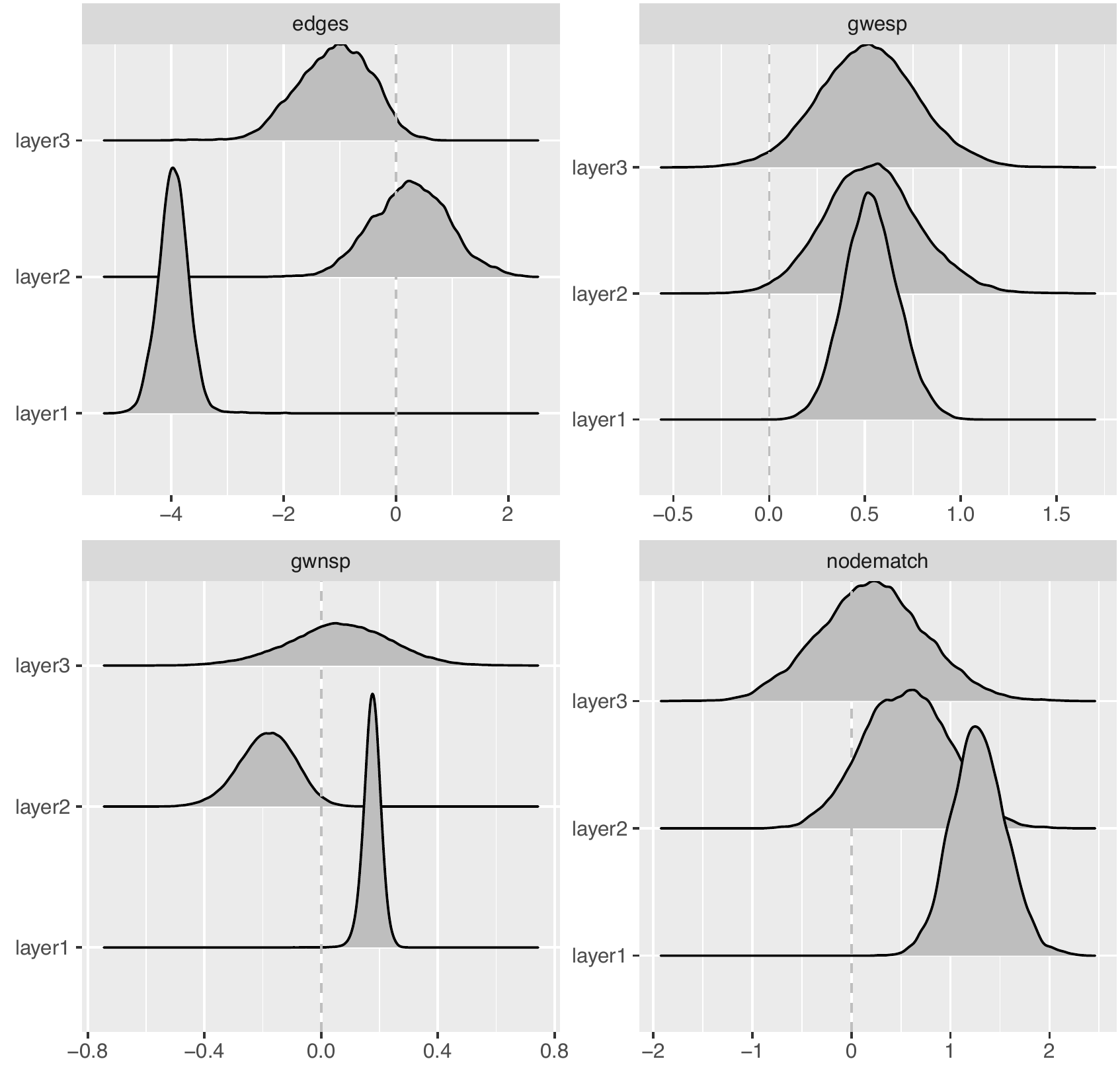}
\caption{Predictive posterior of $\bfphi_{1}$ ({\sf edges}), $\bfphi_{2}$ ({\sf gwesp}), $\bfphi_{3}$ ({\sf gwnsp}), $\bfphi_{4}$ ({\sf nodematch}) for the Zachary karate club network.}
\label{fig:ppPhi_zach1}
\end{figure}

More specifically the estimated predictive posterior for $\bfphi_{1}, \bfphi_{2}, \bfphi_{3}, \bfphi_{4}$, displayed in Figure~\ref{fig:ppPhi_zach1} and summarised in Table~\ref{tab:ppPhi_zach1}, indicates that most of the density dissolution process described above and represented by the parameter $\bfphi_1$ (corresponding to the {\sf edges} statistic) is stronger in the first and between the second and the third network layer. We can also observe the importance of the transitivity effect captured by $\bfphi_2$ (corresponding to the {\sf gwesp} statistic) between all the layers meaning that the dissolution ERGM process across layers affects mainly edges that are not embedded into triadic transitive structures. We can therefore realise that the strongest edges, i.e., the ones surviving the multilayer ERGM dissolution process, are mainly the ones involved in transitive triads. It is interesting to notice how the parameter $\bfphi_3$ (corresponding to the {\sf gwnsp} statistic) contributes positively to the generation of the first layer and negatively to the generation of the second layer conditional on the first layer. This can be explained by the fact that connectivity structure of layer 1 is very much influenced by the presence of open transitive triads involving the two competing leaders of the karate club factions (John and Mr. Hi) who are not directly connected but share many partners among the members of the club. However when we consider the transition from layer 1 to layer 2 many connections involved in these open triads dissolve meaning that they are mostly weak edges. The faction density representing homophily between members of the same faction, captured by $\bfphi_4$ (corresponding to the {\sf nodematch} statistic), is important in explaining the generation of the first layer but not for explaining the dissolution process between the layers.

\begin{table}[htp]
\centering
\caption{Predictive posterior estimates of $\bfphi_{1}, \bfphi_{3}, \bfphi_{3}, \bfphi_{4}$ for the Zachary karate club network.}
\label{tab:ppPhi_zach1}
\begin{tabular}{lcccccc}
                           & \multicolumn{2}{c}{Layer 1} & \multicolumn{2}{c}{Layer 2} & \multicolumn{2}{c}{Layer 3} \\
Parameter (Effect)         & Mean  & SD   & Mean  & SD   & Mean  & SD   \\ \hline
$\bfphi_1$ ({\sf edges})     & -3.96 & 0.27 & 0.27  & 0.68 & -1.09 & 0.63 \\
$\bfphi_2$ ({\sf gwesp})     & 0.53  & 0.14 & 0.53  & 0.24 & 0.51  & 0.25 \\
$\bfphi_3$ ({\sf gwnsp})     & 0.17  & 0.02 & -0.18 & 0.09 & 0.06  & 0.17 \\
$\bfphi_4$ ({\sf nodematch}) & 1.29  & 0.27 & 0.56  & 0.43 & 0.21  & 0.52 \\ \hline
\end{tabular}
\end{table}

\subsection{Model assessment}

A way to examine the fit of the data to the estimated posterior distribution of the parameters is to implement a graphical Bayesian goodness-of-fit procedure \citep{hun:goo:han08}. In the Bayesian context,  network data are simulated from a sample of parameter values drawn from the estimated posterior distribution and compared to the observed data in terms of high-level network characteristics that are not explicitly included as sufficient statistics in the model. Since we are dealing with weighted networks, we focus on the weighted degree distribution. The black solid lines represent the distribution of the weighted degrees in the observed data, the grey lines represent the distribution of the weighted degrees calculated on network graphs simulated from the estimated posterior density. 

\begin{figure}[htp]
\centering
\includegraphics[scale=0.5]{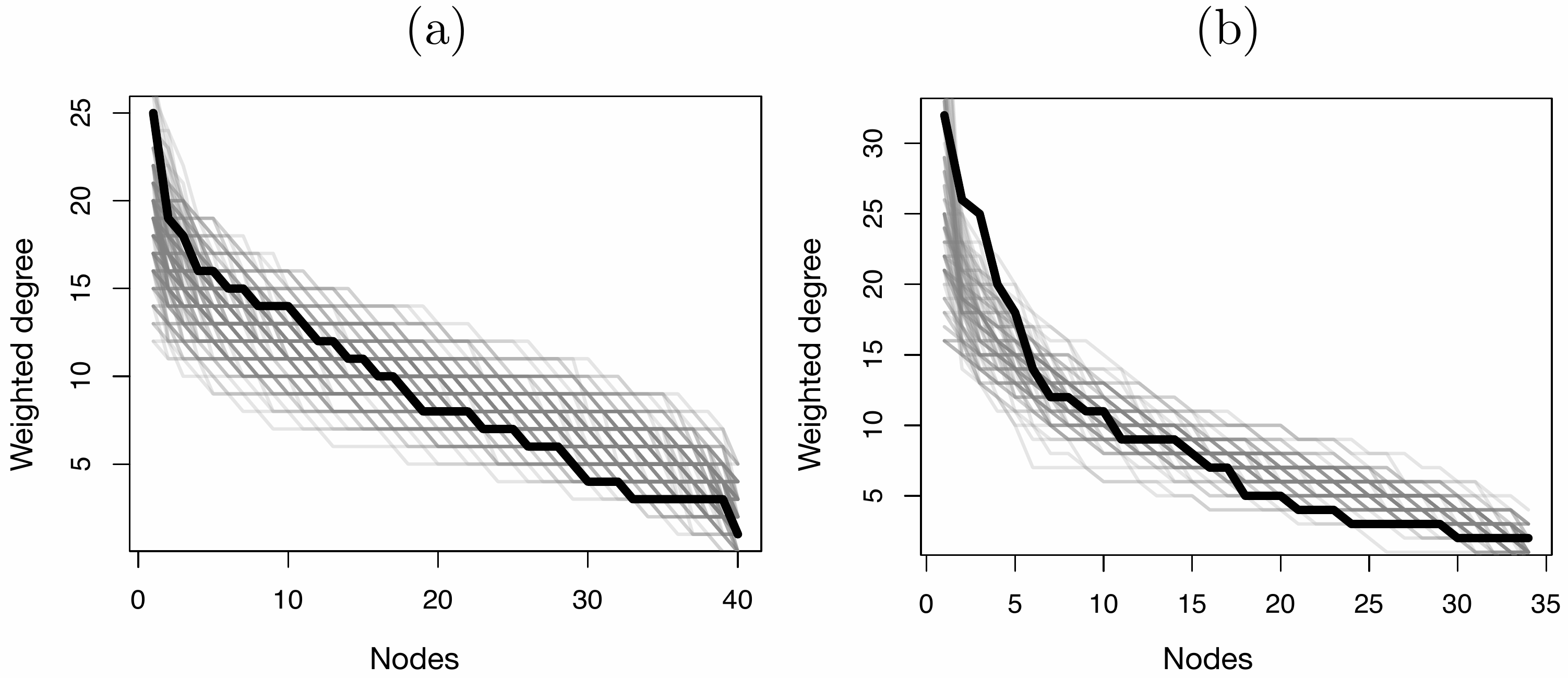}
\caption{Weighted degrees for each node in (a) Bernard and Killworth office network and (b) Zachary karate club network. The black line and the gray lines represent the weighted degree values for the observed network and simulated networks, respectively.}
\label{fig:bgofs}
\end{figure}

The plots in Figure~\ref{fig:bgofs} suggest that both models are a reasonable fit to their respective datasets as the black lines representing the observed weighted degree distributions are lying on the high posterior predictive distribution despite the absence of a degree-based statistic in the Zachary karate club model (this absence is only partially compensated by the presence of the open triadic effect corresponding to the {\sf gwnsp} statistic.)

\section{Conclusions}\label{sec:concl}
This paper has introduced a new Bayesian hierarchical ERGM framework for the analysis of weighted networks with ordinal/polytomous edges which complements the recent advances proposed by \cite{wya:cho:bil10,kri12,des:cra12,kri:but17,wil:etal17}. The modelling approach is based on a flexible multilayer ERGM process able to describe the ERGM dissolution process which leads to the generation of the strength of the network weighted edges. The multilayer ERGM process is parametrised using binary network statistics and it is therefore providing a natural interpretation of the network effects that are assumed to be at the basis of the generative process.

A fully-probabilistic Bayesian approach has been adopted to provide the possibility of specifying prior information of the network effects and analysing their posterior distribution given the observed data. An extension of the approximate exchange algorithm \citep{cai:fri11} has been used in order to carry out inference on the doubly-intractable posterior distribution. Probabilistic goodness-of-fit diagnostics based on the weighted degree distribution have been proposed to assess the models used in two applications on well-known datasets. 

The modelling framework proposed in this paper can be extended to deal with weighted signed networks by, for example, making a distinction between a network process generating the baseline interaction layer encoding the presence or absence of any signed relation between nodes and a joint conditional multilayer ERGM process describing the formation of positive and negative weighted relations between the interacting nodes.

Moreover, the affinity between separable temporal ERGMs and multilayer ERGMs can be exploited to propose a joint modelling approach for temporal weighted networks.

\bibliographystyle{abbrvnat}
\bibliography{myref}

\newpage

\section{Appendix}\label{sec:appendix}


\begin{figure}[htp]
\centering
\includegraphics[scale=0.35]{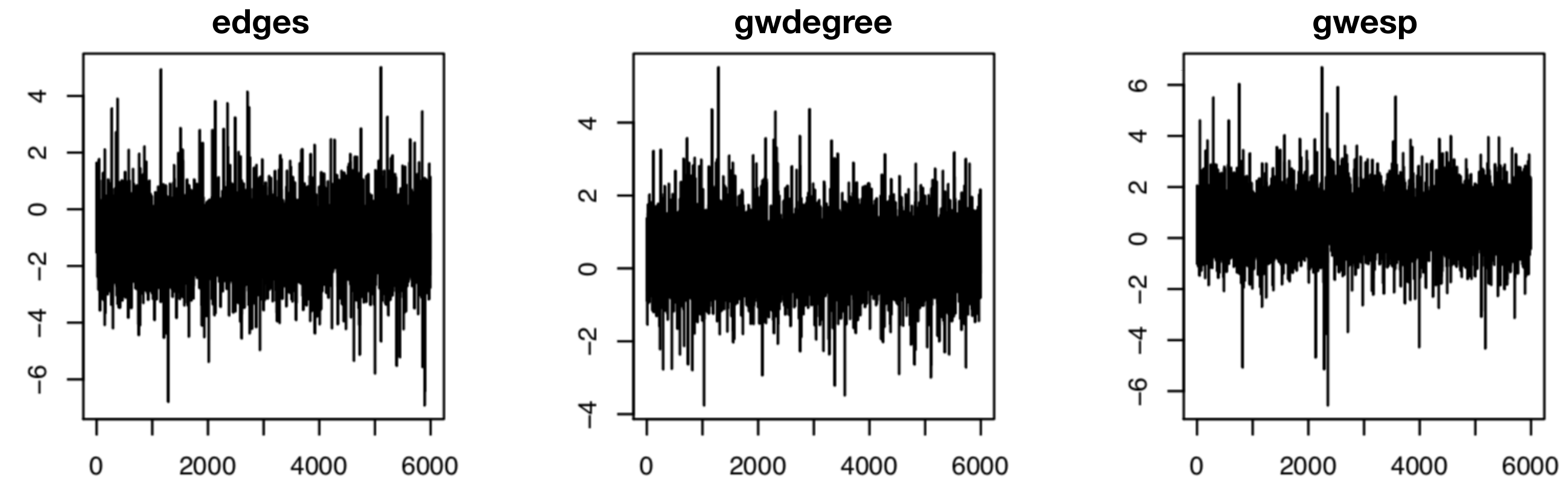}
\caption{Bernard and Killworth office network; MCMC traces for $\bfmu$ (thinning factor = 100).}
\label{fig:phi_traces_bkoff}
\end{figure}

\begin{figure}[htp]
\centering
\includegraphics[scale=0.48]{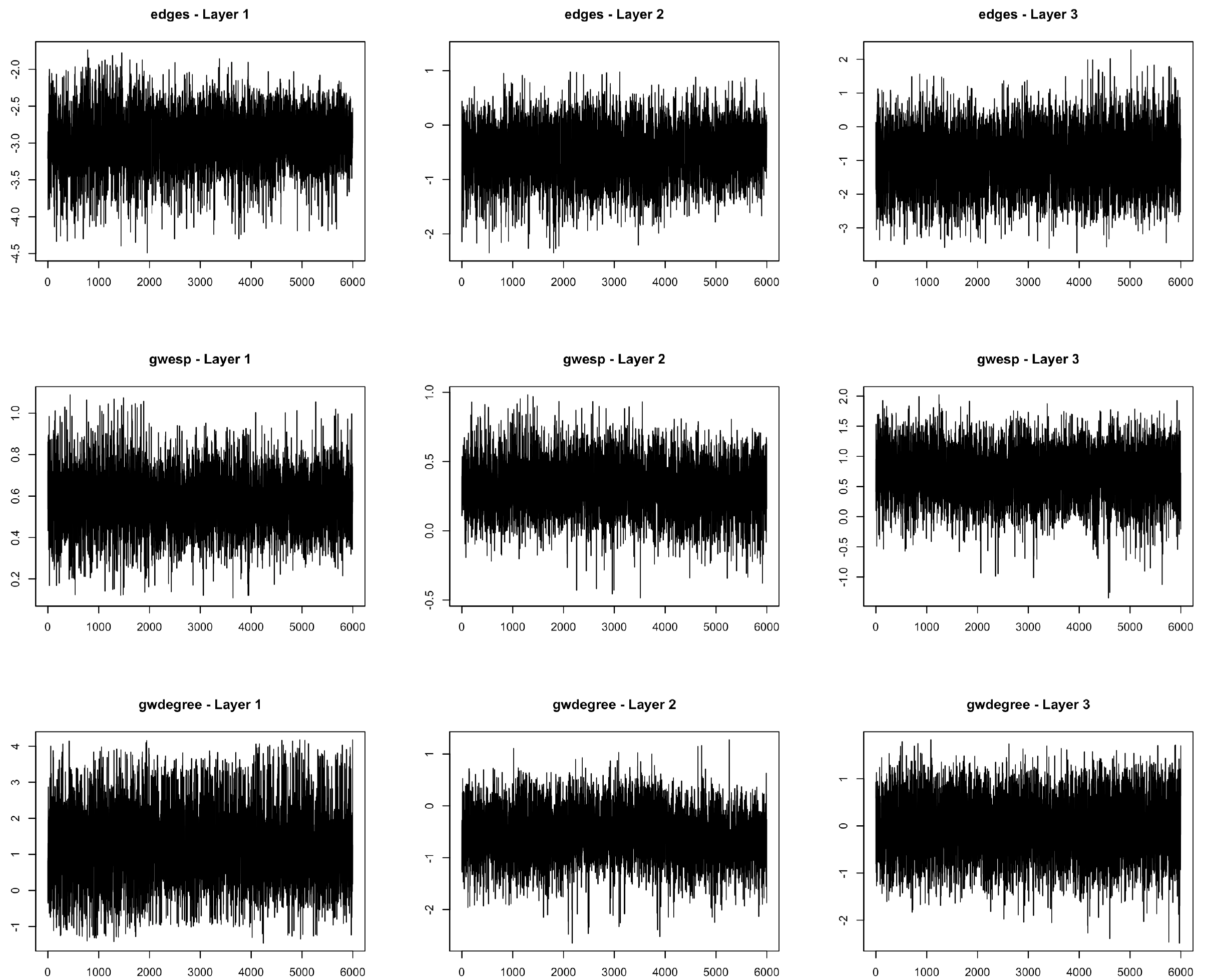}
\caption{Bernard and Killworth office network; MCMC traces for $\bfphi$ (thinning factor = 100).}
\label{fig:phi_traces_bkoff}
\end{figure}

\begin{figure}[htp]
\centering
\includegraphics[scale=0.35]{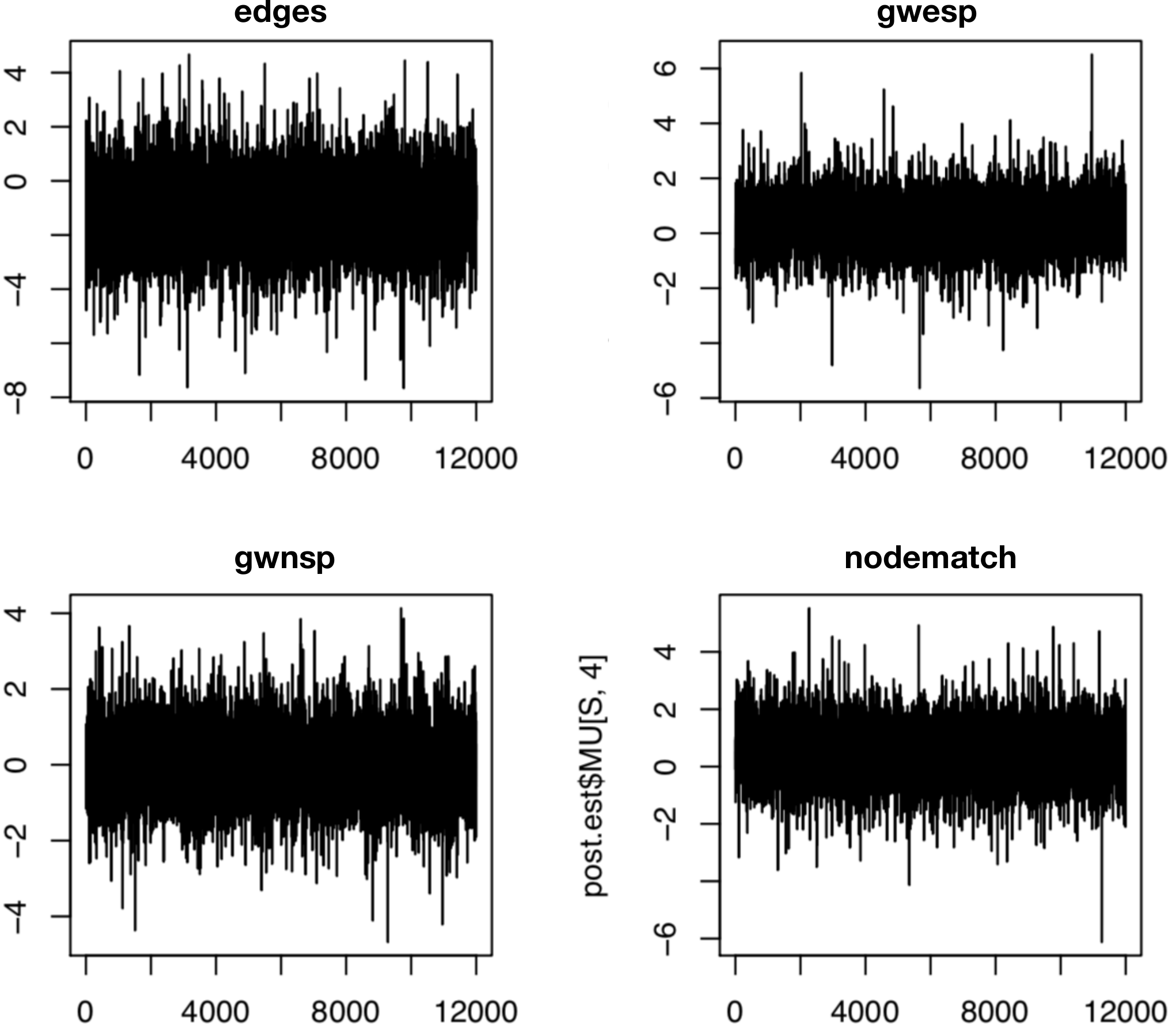}
\caption{Zachary karate club network; MCMC traces for $\bfmu$ (thinning factor = 100).}
\label{fig:phi_traces_bkoff}
\end{figure}

\begin{figure}[htp]
\centering
\includegraphics[scale=0.5]{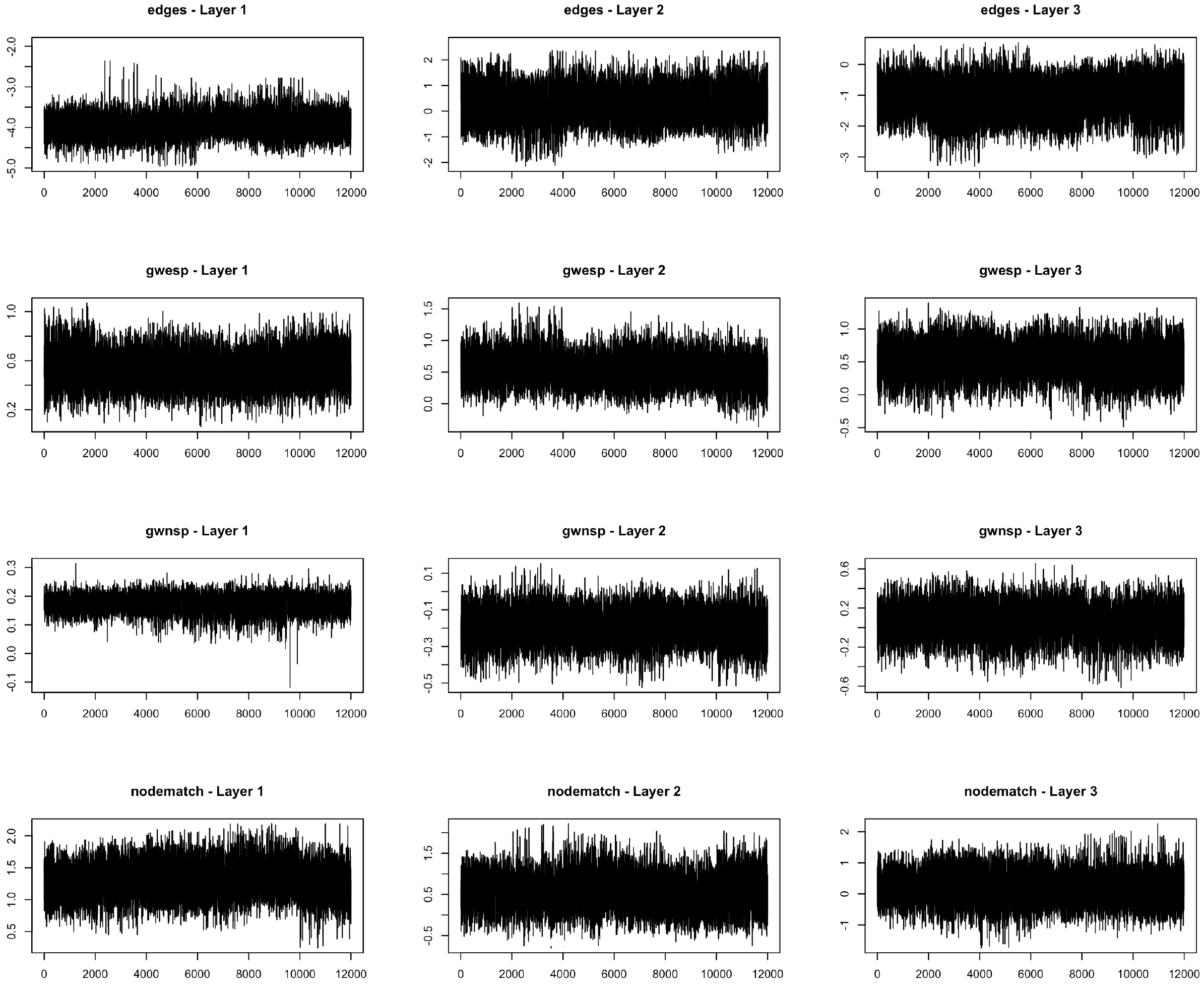}
\caption{Zachary karate club network; MCMC traces for $\bfphi$ (thinning factor = 100).}
\label{fig:phi_traces_zach}
\end{figure}

\end{document}